\documentclass[prd,showkeys,superscriptaddress,
nofootinbib,floatfix,
preprint,
12pt
]{revtex4-2}

\usepackage{booktabs}
\usepackage{epsfig}
\usepackage{tabularx}
\usepackage{multirow}
\usepackage{slashed}
\usepackage{extarrows}
\usepackage{amssymb,amsmath,bm,amsfonts,bbm}
\usepackage{xcolor}
\usepackage{mathtools}
\usepackage{rotating}
\usepackage{booktabs}
\usepackage[colorlinks=true,allcolors=violet,hyperfootnotes=false]{hyperref}
\usepackage{simplewick} 
\usepackage{mathrsfs}

\newcommand{\be}{\begin{equation}} \newcommand{\ee}{\end{equation}}
\newcommand{\ba}{\left(\begin{array}{c}}
\newcommand{\ea}{\end{array}\right)}
\newcommand{\bea}{\begin{eqnarray}} \newcommand{\eea}{\end{eqnarray}}

\newcommand{\rd}{{\rm d}}

\newcommand{\RNum}[1]{\uppercase\expandafter{\romannumeral #1\relax}}

\newcommand{\mL}{\mathcal{L}}

\newcommand{\mD}{\mathcal{D}}
\newcommand{\mP}{\mathcal{P}}

\allowdisplaybreaks[4]

\begin{document}

\title{\Large The $D_{s1}(2460)$ and other open-charm $1^+$ states in relativistic chiral effective field theory}

\author{Ze-Rui~Liang}
\affiliation{College of Physics and Hebei Key Laboratory of Photophysics Research and Application and Physics Postdoctoral Research Station at Hebei Normal University, Hebei Normal University, Shijiazhuang, Hebei 050024, China}
\affiliation{School of Physics and Electronics, Hunan University, 410082 Changsha, China}

\author{Qi-Chao~Xiao}
\affiliation{School of Physics and Electronics, Hunan University, 410082 Changsha, China}
\affiliation{Hunan Provincial Key Laboratory of High-Energy Scale Physics and Applications, Hunan University, 410082 Changsha, China}

\author{Zhi-Hui Guo}
\affiliation{College of Physics and Hebei Key Laboratory of Photophysics Research and Application, Hebei Normal University, Shijiazhuang, Hebei 050024, China}

\author{De-Liang~Yao}
\email{yaodeliang@hnu.edu.cn}
\affiliation{School of Physics and Electronics, Hunan University, 410082 Changsha, China}
\affiliation{Hunan Provincial Key Laboratory of High-Energy Scale Physics and Applications, Hunan University, 410082 Changsha, China}

\begin{abstract}
We derive the pertinent chiral potentials for charmed vector meson interactions with light pseudoscalar bosons in a relativistic U(3) chiral effective field theory up to next-to-leading order. Predictions for the $S$- and $P$-wave scattering lengths are obtained for all the relevant elastic channels. A comparison with the most recent---and currently the sole---lattice QCD data on the $S$-wave $I=1/2$ $D^\ast\pi$ scattering length at a pion mass of $391$~MeV reveals good agreement, thereby validating the estimation of low energy constants via heavy quark spin symmetry. Within the relativistic formalism, we confirm that the $D_{s1}(2460)$ can be identified with a bound state pole, while the $D_1(2430)$ corresponds to the interplay of two poles: a lower one on the second Riemann sheet and a higher one on the third Riemann sheet. We show that the $D_{s1}(2460)$ and the lower $D_1(2430)$ pole originate from the same flavor SU(3) triplet, whereas the higher $D_1(2430)$ pole belongs to the SU(3) sextet. All these states are not of $\bar{q}q$ nature, as they flow to complex infinity in the large $N_C$ limit. Our results provide quantitative benchmarks for future lattice QCD and femtoscopic studies.

\end{abstract}

\date{\today}
\maketitle

%
%
\newpage
\section{Introduction}

The discovery of the charm-strange scalar ($J^P=0^+$) and axial-vector ($1^+$) mesons $D_{s_0}^\ast(2317)$~\cite{BaBar:2003oey} and $D_{s1}(2460)$~\cite{CLEO:2003ggt} has triggered intensive studies of hadronic interactions involving charmed mesons, owing to their masses being significantly lower than the quark model expectation~\cite{Godfrey:1985xj,Godfrey:1986wj}. A variety of interpretations have been proposed to explain the nature of these two states including $c\bar{s}$ state with quenching effect, compact tetraquark and hadronic molecule; see. e.g., Refs~\cite{Chen:2016spr, Guo:2017jvc, Meng:2022ozq, Liu:2024uxn, Wang:2025sic} for review. The subsequent observation of $D_0^\ast(2300)$~\cite{FOCUS:2003gru} and $D_1(2430)$~\cite{Belle:2003nsh} has made the classification of open-charm states even more puzzling, as their masses are comparable to those of their charm-strange counterparts, respectively. In addition, the description of the broad $D_0^\ast(2300)$ and $D_1(2430)$ resonances is challenging, like the $f_0(500)$ and $K^\ast_0(700)$ in the light scalar meson sector~\cite{Pelaez:2015qba, Yao:2020bxx}. For instance, the standard Breit-Wigner (BW) parametrization~\cite{Du:2020pui}, valid only for narrow resonances yet commonly used in experimental analyses, is inadequate for broad resonance due to unavoidable coupled-channel effects. Moreover, it fails to satisfy the chiral symmetry constraints that the amplitudes for reactions involving these resonances must respect~\cite{Du:2019oki}. A natural framework for describing the above open-charm excited excitations is provided by chiral perturbation theory (ChPT) for heavy mesons~\cite{Burdman:1992gh, Yan:1992gz, Wise:1992hn} together with unitarization approaches~\cite{Gasser:1990bv, Dobado:1996ps, Oller:1997ti,Oller:1998zr} (see, e.g., Ref.~\cite{Oller:2019opk, Yao:2020bxx} for reviews).

Chiral perturbation theory~\cite{Weinberg:1978kz, Gasser:1983yg, Gasser:1984gg}, as the effective field theory of Quantum Chromodynamics (QCD) at low energies, is a powerful tool for investigating the non-perturbative strong interactions with pseudoscalar Nambu Goldstone bosons (pNGBs, $\phi\in\{\pi, K, \bar{K}, \eta\}$) in a systematically controlled manner. It was demonstrated in Refs.~\cite{Kolomeitsev:2003ac, Guo:2006fu,Gamermann:2006nm,Altenbuchinger:2013vwa,Altenbuchinger:2013gaa,Guo:2015dha,Albaladejo:2016lbb, Du:2017zvv, Luo:2026kui,Zhuang:2026lta} that its unitarized version, known as unitarized ChPT (UChPT), does offer a compelling resolution to the mass-ordering puzzle of the lightest strange and non-strange scalars, as well as axial vectors. Moreover, the $D_0^\ast(2300)$ can be naturally interpreted as the interplay of two dynamically generated poles~\cite{Albaladejo:2016lbb,Zhuang:2026lta,Luo:2026kui}; see also Ref.~\cite{Meissner:2020khl} for review of two-pole structures in QCD. The success of UChPT in this context stems not only from its automatic preservation of QCD symmetries, such as chiral symmetry, but also from its systematic incorporation of coupled-channel effects through chiral potentials. In addition, unitarity can be conveniently realized by various unitarization methods, which take the chiral potentials as input kernel. Thus, chiral potential is at the heart of UChPT. For scatterings of pNGBs off charmed pseudoscalar mesons ($\mathcal{P}\in\{D,D_s\}$), chiral potentials have been derived at tree~\cite{Kolomeitsev:2003ac,Guo:2006fu,Gamermann:2006nm, Guo:2009ct,Liu:2012zya, Altenbuchinger:2013gaa, Altenbuchinger:2013vwa} and one-loop level~\cite{Liu:2009uz,Geng:2010vw,Yao:2015qia, Du:2017ttu, Huang:2021fdt,Lutz:2022enz,Isken:2023xfo}, based on the chiral effective Lagrangians constructed, e.g., in Ref.~\cite{Jiang:2019hgs}. The low energy constants (LECs) involved in the $\mathcal{P}\phi$ potentials can be well determined by fitting to lattice QCD data accumulated in the recent years~\cite{Liu:2012zya,Mohler:2013rwa,Mohler:2012na,Moir:2016srx,Bali:2017pdv,Cheung:2020mql,Yan:2024yuq}. Positivity bounds are also derived for the $\mathcal{P}\phi$ LECs with the help of axiomatic $S$ matrix such as unitarity, analyticity and crossing symmetry~\cite{Du:2016tgp}. The well-determined LECs enable one to make reliable predictions of various physical observables for other decay processes via final state interactions~\cite{Du:2017zvv,Yao:2018tqn}.

So far, the potentials for interactions between charmed vector mesons ($\mathcal{P}^\ast\in\{D^\ast,D^\ast_s\}$) and pNGBs are less developed. Though the Weinberg-Tomazawa (WT) term of leading order (LO) is entirely fixed by chiral symmetry, the currently available lattice QCD data~\cite{Lang:2022elg,Lang:2025pjq} are rare and insufficient to pin down the relevant LECs appearing in $\mathcal{P}^\ast\phi$ scatterings at next-to-leading order (NLO) and beyond. Instead, the so-called heavy quark spin symmetry (HQSS)~\cite{Isgur:1991wq} is usually used to estimate the $\mathcal{P}^\ast\phi$ LECs from those of $\mathcal{P}\phi$ interactions. Nevertheless, the validity of HQSS-based estimation of the LECs in $\mathcal{P}^\ast\phi$ scatterings has not yet been thoroughly tested. Given the recent emergence of both lattice QCD~\cite{Lang:2022elg,Lang:2025pjq} and experimental~\cite{ALICE:2024bhk} data concerning the $\mathcal{P}^\ast\phi$ interactions, it is now timely to make such a justification. Specifically, on the one hand, the hadron spectrum collaboration (HSC) has recently performed the extractions of charmed axial-vector mesons in the $I=1/2$ $D^\ast\pi$~\cite{Lang:2022elg} and $D^\ast\pi$-$D^\ast\eta$-$D_s^\ast\bar{K}$~\cite{Lang:2025pjq} amplitudes from lattice simulations at a pion mass of $391$~MeV, from which the $S$-wave scattering lengths can be readily obtained. On the other hand, by using femtoscopy technique~\cite{Pratt:1986cc} (see also Ref.~\cite{Liu:2024uxn} for a comprehensive review), the first measurement of the residual strong $\mathcal{P}^{(\ast)}\phi$ interactions has been conducted by ALICE collaboration~\cite{ALICE:2024bhk}. The $I=1/2$ and $I=3/2$ $D^{(\ast)}\pi$ scattering lengths are extracted from the measured two-particle correlation functions. Unexpectedly, tensions have been found between these experimental extractions with predictions from both ChPT and lattice QCD calculations. One of the primary objectives of this work is to test the HQSS-based predictions against the HSC data within U(3) ChPT for the $\mathcal{P}^\ast\phi$ system,  and to examine whether the tension persists when the NLO $\mathcal{P}^\ast\phi$ potentials are utilized as the unitarization kernel. 

In this work, we derive the $\mathcal{P}^\ast\phi$ scattering amplitudes in a relativistic U(3) ChPT up to NLO. The $\eta_0$ is included as an active U(1) singlet. Compared to the usual SU(3) framework with $\pi$, $K$, $\bar{K}$ and $\eta_8$ used in Refs.~\cite{Kolomeitsev:2003ac, Hofmann:2003je, Guo:2006rp,Gamermann:2007fi, Guo:2018gyd}, the inclusion of $\eta_0$ yields a more realistic description of the physical $\eta$ and $\eta^\prime$ states, since they are admixture of $\eta_8$ and $\eta_0$. It also allows us to explore the large-$N_C$ (number of colors in QCD) behavior of dynamically generated poles, since the $\eta_0$ corresponds to the ninth pNGB in the large $N_C$ limit~\cite{Witten:1979vv,Veneziano:1979ec,Coleman:1980mx,Manohar:1998xv}. The obtained interaction potentials consist of the WT term, NLO contact terms as well as explicit $D_{(s)}^{(\ast)}$-meson exchanges in the $s$- and $u$-channels. After a relativistic partial-wave decomposition in the helicity basis, the amplitudes are unitarized with the on-shell Bethe--Salpeter equation (BSE)~\cite{Oller:2000fj} to restore unitarity above thresholds. The NLO LECs $h_i^\ast$ ($i=0,2,3,4,5$), which can not be fixed by experimental inputs, are set equal to their corresponding well-established $\mathcal{P}\phi$ LECs~\cite{Guo:2015dha} in accordance with HQSS. The formalism is applied to four elastic and three coupled channels, classified by strangeness ($S$) and isospin ($I$). We calculate elastic $S$- and $P$-wave scattering lengths and their pion-mass dependence, and compare them with HSC lattice extractions~\cite{Lang:2025pjq} and ALICE femtoscopic measurements~\cite{ALICE:2024bhk}. Our prediction of the $S$-wave $D^\ast\pi$ ($I=1/2$) scattering length, calculated at unphysical pion mass $M_\pi=391$~MeV, is in remarkable agreement with the HSC result, strongly supporting the validity of HQSS. However, as for the $I=1/2$ and $I=3/2$ $D^\ast\pi$ scattering lengths, tensions persist between the ALICE extractions and our determinations with NLO potentials. The meson-exchange contribution is found to be negligible for the $S$-wave scattering lengths, yet it exerts a significant influence on the $P$-wave scattering lengths in the $(S,I)=(1,0)$ and $(0,1/2)$ coupled channels. This is attributed to the presence of the $s$-channel meson exchange, which is allowed exclusively in these channels by the conservation of angular momentum ($J$) and parity ($P$). 

We search for dynamically generated $1^+$ poles in the unitarized amplitudes on different Riemann sheets (RSs) of complex $s$ plane. Specifically, a bound-state pole is found at $\sqrt{s_{\rm pole}}=2455.2^{+3.2}_{-2.7}$~MeV, located below the $D^\ast K$ threshold, which can be identified with the experimentally observed $D_{s1}(2460)$ state~\cite{ParticleDataGroup:2026aaa}. Furthermore, a two-pole picture of the iso-doublet axial-vector $D_1(2430)$~\cite{ParticleDataGroup:2026aaa}, the HQSS partner of $D_0^\ast(2300)$, is likewise established. Namely, the $D_1(2430)$ resonance corresponds to the superposition of two poles: a lower pole on the $2$nd RS at $\sqrt{s_{\rm pole}}=2255.6^{+3.3}_{-2.8}-i\,112.5_{-2.9}^{+2.6}$~MeV, and a higher one on the $3$rd RS at $\sqrt{s_{\rm pole}}=2558.1^{+31.0}_{-23.4}-i\, 207.2^{+7.8}_{-8.4} $~MeV. The properties of these poles are further explored by tracing their trajectories as functions of $M_\pi$, $N_C$ and the parameter $x$, which characterizes the deviation from the flavor SU(3)-symmetric limit toward the physical case. Our primary findings are as follows. The evolution with $x$ reveals that the $D_{s1}(2460)$ and the lower $D_1(2430)$ pole originate from the same flavor SU(3) triplet, namely the $\bar{\mathbf{3}}$ irreducible representation ({\it irrep}), while the higher $D_1(2430)$ pole arises from ${\mathbf{6}}$ {\it irrep}. Upon increasing $M_\pi$ up to $670$~MeV, the $D_{s1}(2460)$ and the higher $D_1(2430)$ poles retain their characters as a bound-state and a resonance, respectively. In contrast, the lower $D_1(2430)$ pole involves from a resonance to a bound state, as a consequence of its steadily increasing coupling strength to the $D^\ast\pi$ channel. The existence of a $D_1$ bound-state pole at large $M_\pi$ is supported by the recent HSC result~\cite{Lang:2025pjq}. Finally, all these poles flow to complex infinity as $N_C\to\infty$, indicating that they are not of $\bar{q}q$ nature.

The manuscript is organized as follows. In Sec.~\ref{sec:formalism} we introduce the U(3) chiral Lagrangians and derive the $\mathcal{P}^{\ast}\phi$ potentials, partial-wave amplitudes, and unitarized scattering amplitudes. Numerical inputs, scattering lengths, and $S$-matrix parameters are presented in Sec.~\ref{sec:numerical}. Section~\ref{sec:pole_trajectories} is devoted to pole extractions, SU(3) analysis, and pole trajectories with varying $M_\pi$ and $N_C$. We summarize in Sec.~\ref{sec:summary}. Explicit expressions for some of the coefficients appearing in the chiral potentials and for the Lorentz invariant amplitudes are relegated to Appendices~\ref{Apend:Co} and~\ref{Apend:amp}, respectively.

\section{Theoretical formalism}\label{sec:formalism}
\subsection{Chiral effective Lagrangians}
The chiral effective Lagrangian relevant to our calculation up to NLO can be organized as~\cite{Wise:1992hn,Burdman:1992gh,Yan:1992gz}
\begin{align}
\label{Chlag}
\mL_{\rm eff}= \mathcal{L}^{(1)}_{\mP^\ast\phi}+\mathcal{L}^{(1)}_{\mP\phi}+\mathcal{L}^{(1)}_{\mP^\ast\mP\phi}+\mathcal{L}^{(1)}_{\mP^\ast\mP^\ast\phi}+\mathcal{L}^{(2)}_{\mP^\ast\phi} \ ,
\end{align} 
with the numbers in the superscripts denoting the chiral orders. The first two pieces are kinematic terms, which read
\begin{align}
\label{lolag:ct}
&\mathcal{L}^{(1)}_{\mP^\ast\phi}=-\mathcal{D}_\mu \mP^\ast_\alpha \mathcal{D}^\mu \mP^{\ast\alpha\dagger}+\overline{M}_{\mP^\ast}^2 \mP_\alpha^\ast \mP^{\ast\alpha\dagger}\ , \\
&\mathcal{L}^{(1)}_{\mP\phi}=\mathcal{D}_\mu \mP \mathcal{D}^\mu \mP^{\dagger}-\overline{M}_{\mP}^2 \mP \mP^{\dagger}\ , 
\end{align} 
where $\overline{M}_{{\mP}^\ast}$ and $\overline{M}_{\mP}$ are the masses of the charmed vector mesons $\mP^\ast=(D^{\ast0},D^{\ast+},D^{\ast+}_s)$ and pseudo-scalar charmed mesons $\mP=(D^0,D^+,D_s^+)$ in the chiral limit, respectively. The covariant derivative $\mathcal{D}_\mu$ acting on the charmed meson fields is defined by
\begin{align}
\mathcal{D}_\mu X= X(\overset{\leftarrow}{\partial_\mu}+\Gamma_\mu^\dagger)\ ,\quad
\mathcal{D}_\mu X^\dagger=(\partial_\mu+\Gamma_\mu) X^\dagger\ ,
\end{align} 
where $X\in\{\mP,\mP^\ast\}$ and the chiral connection is defined by $\Gamma_\mu = \frac{1}{2}[u^\dagger\partial_\mu u+u\partial_\mu u^\dagger]$. In U(3) framework, the pNGBs are non-linear parametrized in the matrix $U$, 
\begin{align}
U&=u^2= {\rm exp}( i{\sqrt{2}\phi}/{F_{0}})\ ,\quad
\phi= \begin{pmatrix}
\frac{\pi^0}{\sqrt{2}}+\frac{\eta_8}{\sqrt{6}}   &\pi^+                                                                  &K^+\\
\pi^-                                                                  &-\frac{\pi^0}{\sqrt{2}}+\frac{\eta_8}{\sqrt{6}}   &K^0 \\
K^-                                                                    &\bar{K}^0                                                           &-\frac{2\eta_8}{\sqrt{6}}
\end{pmatrix} +\frac{\mathbbm{1}}{\sqrt{3}}\eta_0\ .\label{eq.U}
\end{align}
 Here the parameter $F_0$ is the axial decay constant of the pNGBs in the chiral limit. The $U$ matrix has been extended to an U(3) version by adding the term proportional to the singlet $\eta_0$. The reason for this extension is twofold. One the one hand, it completes the low-energy QCD spectrum in the large $N_C$ limit, wherein the singlet $\eta_0$ emerges as a relevant degree of freedom. On the other hand, it enables a more realistic description of the physical $\eta$ state, which arises from the mixing between $\eta_8$ and $\eta_0$, as does the $\eta^\prime$. Furthermore, the QCD U(1)$_A$ anomaly, scaling as $1/N_C$~\cite{Witten:1979vv,Veneziano:1979ec,Coleman:1980mx,Manohar:1998xv}, is responsible for the massive $\eta_0$ in the physical situation, which tends to massless as $N_C\to\infty$ and chiral limit. Therefore, in reality, the LO chiral Lagrangian for the light pNGBs should be generalized to
\begin{align}
\label{eq:Lphi}
\mathcal{L}_\chi
= \frac{F^2}{4}\langle u_\mu u^\mu\rangle
+ \frac{F^2}{4}\langle \chi_+\rangle
+ \frac{F^2}{12} M_0^2\, X^2\,,
\quad X=\ln(\det U)=\sqrt{6}\frac{i}{F}\eta_0+\mathcal{O}(\eta_0^3)\,,
\end{align}
where the symbol $\langle\cdots\rangle$ represents the trace in the flavour space. The definition of $\chi_{+}$ is given by $\chi_+=u^\dagger \chi u^\dagger+u \chi^\dagger u$, with $\chi= 2B {\rm diag} \{ \hat{m},\hat{m},m_s \} $, where $\hat{m}$ is the average mass of the light $u$ and $d$ quarks, and $m_s$ denotes the mass of strange quark. The factor $B$ is a constant related to the quark condensate. $M_0$ is the LO mass of $\eta_0$, which is suppressed by $1/N_C$ in QCD but kept finite in the U(3) construction. The physical $\eta$ and $\eta'$ are related to the singlet-octet basis $(\eta_8,\eta_0)$ by an
orthogonal rotation,
\begin{align}
\label{eq:mix}
\begin{pmatrix} \eta \\ \eta' \end{pmatrix}
=
\begin{pmatrix}
\cos\theta & -\sin\theta \\
\sin\theta & \phantom{-}\cos\theta
\end{pmatrix}
\begin{pmatrix} \eta_8 \\ \eta_0 \end{pmatrix}\,,
\end{align}
with the mixing angle $\theta$. The masses and mixing angle are given by
\begin{align}
   M_{\eta^\prime/\eta}^2 = M_K^2 + \frac{M_0^2}{2} \pm \frac{R}{2}\,,\quad
   \sin\theta=-\bigg(\sqrt{1+\frac{S^2}{32\Delta^4}}\bigg)^{-1}\,,\label{eq.eta.etap}
\end{align}
where $3R \equiv \sqrt{9M_0^4 - 3M_0^2\Delta^2 + 12\Delta^4}$ and $S\equiv 3M_0^2-2\Delta^2+3R$ with $\Delta\equiv M_K^2-M_\pi^2$. 
In the limit $M_0\to \infty$, one has $\sin\theta\to 0$ and $\cos\theta\to1$. Thus, the $\eta_{0,8}$ decouple according to Eq.~\eqref{eq:mix}. In this limit, the physical $\eta$ can be identified as the octet $\eta_8$ in the SU(3) ChPT~\cite{Gasser:1984gg}, and its mass reduces to
\begin{align}
M_\eta^2 &= \frac{M_0^2}{2}+M_K^2-\frac{R}{2}
= M_K^2+\frac{\Delta^2}{3}
+\mathcal{O}\!\left(\frac{\Delta^4}{M_0^2}\right)
= \frac{4M_K^2-M_\pi^2}{3}
\,,
\label{eq:MetaLargeM0}
\end{align}
in align with the Gell-Mann-Okubo relation~\cite{Gell-Mann:1968hlm}. For completeness, we also derive the LO pion and kaon mass from Eq.~\eqref{eq:Lphi}, which are
\begin{align}
    M_\pi^2=2 B\hat{m}\ ,\quad M_K^2=B(\hat{m}+m_s)\ .\label{eq.pi.K.mass}
\end{align}

The LO $\mathcal{P}^\ast\mathcal{P}\phi$ and $\mathcal{P}^\ast\mathcal{P}^\ast\phi$ interactions are described by
\begin{align} \label{lolag:exc}
&\mathcal{L}^{(1)}_{\mP^\ast\mP\phi}=i g_0\left(\mP^\ast_{\mu}u^\mu \mP^\dagger-\mP\,u^\mu \mP^{\ast\dagger}_\mu\right)\ ,  \\
&\mathcal{L}^{(1)}_{\mP^\ast\mP^\ast\phi}=\dfrac{g_{1}}{2} \left(\mP^\ast_\alpha u_\mu\mD_\nu\mP^{\ast\dagger}_\beta-\mD_\nu\mP^\ast_\alpha u_\mu \mP^{\ast\dagger}_\beta\right)\epsilon^{\alpha\beta\mu\nu}\ ,
\end{align}
where $g_0$ and $g_1$ are coupling constants and $\epsilon^{\alpha\beta\mu\nu}$ is the Levi-Civita tensor. The building block $u_{\mu}$ has the form $u_{\mu}=i [u^\dagger \partial_\mu u-u\partial_\mu u^\dagger ]$.

The Lagrangian at NLO reads~\cite{Guo:2009ct}
\begin{align}\label{nlolag}
\mathcal{L}^{(2)}_{\mP^\ast\phi} &= -\mP^\ast_{\alpha} \big(-{h}^\ast_0\langle\chi_+\rangle-h^\ast_1{\chi}_+
\big) {\mP}^{\ast\alpha\dag} -\mP^\ast_{\alpha} \big(
 h^\ast_2\langle u_\mu u^\mu\rangle
-h^\ast_3u_\mu u^\mu\big) {\mP}^{\ast\alpha\dag} \notag\\
&- \mathcal{D}_\mu \mP^\ast_\alpha\left({h^\ast_4}\langle u_\mu
u^\nu\rangle-{h^\ast_5}\{u^\mu,u^\nu\}\right)\mathcal{D}_\nu {\mP}^{\ast\alpha\dag}\ ,
\end{align}
where $h^{\ast}_i(i=0,1,...,5)$ are unknown LECs, which are usually determined by experimental results or lattice data. These operators yield NLO corrections to the charmed vector meson masses and the contact $\mathcal{P}^\ast\phi$ interactions.

\subsection{Potentials}
\begin{figure}[t]
\centering
\includegraphics[width=0.575\linewidth]{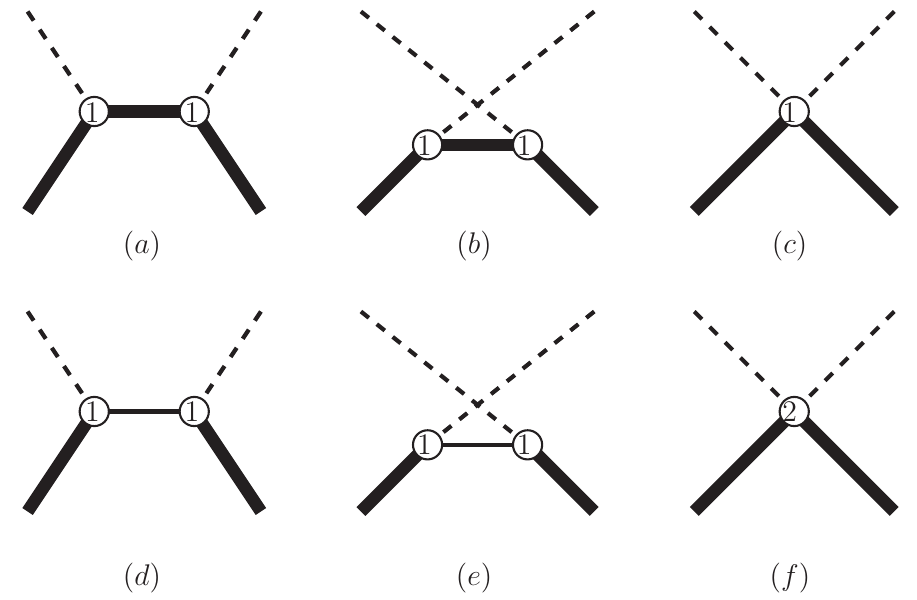} 
\caption{$\mP^\ast\phi$ scattering up to NLO. The thick solid, dashed and thin solid lines represent the $\mathcal{P}^\ast$ mesons, pNGBs and $\mathcal{P}$ mesons, in order. The numbers in the circles indicate the chiral dimensions of the vertices.}
\label{fig:2}
\end{figure}

\begin{table*}[hbp]
\centering
\large
\renewcommand{\arraystretch}{1.2}
\caption{\label{tab:ciex}The coefficients involved in the chiral U(3) potentials~\eqref{eq.pot.ct} and~\eqref{eq.pot.ex}. In the table, we use the abbreviations: $c_\theta=\cos \theta$ and $s_\theta=\sin \theta$. The chiral SU(3) potentials can be obtained as a limiting case by taking $c_\theta=1$, $s_\theta=0$, and switching off all channels that involve the $\eta^\prime$ meson.  Furthermore, some coefficients, e.g., $\mathcal{C}_1^{(1,0)\,K\eta}$, are not shown explicitly; their explicit expressions can be found in Appendix~\ref{Apend:Co}.}
\resizebox{\textwidth}{!}{
\begin{tabular}{|l | c | c c c c c c c | }
\noalign{\global\arrayrulewidth=2pt}
\hline
\noalign{\global\arrayrulewidth=0.4pt}
$(S,I)$               &Channels                                                          &$\mathcal{C}_{\text {LO}}$    &$\mathcal{C}_S$           &$\mathcal{C}_U$         &$\mathcal{C}_0$     &$\mathcal{C}_1$        &$\mathcal{C}_{24}$              &$\mathcal{C}_{35}$
\\
\hline
$(-1,0)$            &$D^\ast \bar{K}\to D^\ast \bar K$                       &$-1$                              &$0$                                &$-6$                           &$M_K^2$ &$M_K^2$ &$1$ &$-1$
\\
$(-1,1)$            &$D^\ast \bar{K}\to D^\ast \bar K$                       &$1$                                &$0$                                &$6$                            &$M_K^2$ &$-M_K^2$ &$1$ &$1$
\\
$(2,\frac12)$    &$D^\ast_s K\to D^\ast_s K$                                &$1$                               &$0$                                 &$6$                             &$M_K^2$ &$-M_K^2$ &$1$ &$1$
\\
$(0,\frac32)$    &$D^\ast \pi\to D^\ast \pi$                                     &$1$                               &$0$                                 &$6$                            &$M_\pi^2$ &$-M_\pi^2$ &$1$ &$1$
\\
\hline
$(1,1)$             &$D^\ast_s\pi\to D^\ast_s\pi$                                 &$0$                             &$0$                                  &$0$                            &$M_\pi^2$ &$0$ &$1$ &$0$     
\\
                        &$D^\ast K\to D^\ast K$                                          &$0$                             &$0$                                   &$0$   &$M_K^2$  &$0$ &$1$ &$0$    
\\
                        &$D^\ast K\to D^\ast_s\pi$                                      &$1$                             &$0$                                                     &$6$   &$0$ &$-\frac{M^2_K+M^2_\pi}{2}$ &$0$ &$1$
\\
\hline
$(1,0)$             &$D^\ast K\to D^\ast K$                                         &$-2$                            &$12$                                                   &$0$ &$M_K^2$ &$-2M_K^2$ &$1$ &$2$
\\
                        &$D^\ast K\to D^\ast_s\eta$                                  &$-\sqrt{3}c_\theta$       &$2\sqrt{6}(\sqrt{2}c_\theta+s_\theta)$ &$-2\sqrt{3}(c_{\theta}-\sqrt{2}s_{\theta})$  &$0$ &$\mathcal{C}^{(1,0) \ K\eta}_1$ &$0$ &$\frac{c_\theta+2\sqrt{2} s_{\theta}}{\sqrt{3}}$
\\
                        &$D^\ast_s\eta\to D^\ast_s\eta$                            &$0$                              &$2(\sqrt{2}c_\theta+s_\theta)^2$   &$2(\sqrt{2}c_\theta+s_\theta)^2$ &$\mathcal{C}^{(1,0)\ \eta\eta}_0$ &$\mathcal{C}^{(1,0)\ \eta\eta}_1$ &$1$ &$\frac{2(\sqrt{2}c_\theta+s_\theta)^2}{3}$
\\
                        &$D^\ast K\to D^\ast_s\eta^\prime$                       &$-\sqrt{3}s_\theta$      &$-2\sqrt{6}(c_\theta-\sqrt{2}s_\theta)$   &$-2\sqrt{3}(\sqrt{2}c_\theta+s_\theta)$ &$0$ &$\mathcal{C}^{(1,0)\ K\eta^\prime}_1$ &$0$ &$\mathcal{C}^{(1,0)\ K\eta^\prime}_{35}$
\\
                        &$D^\ast_s\eta\to D^\ast_s\eta^\prime$                 &$0$                              &$\mathcal{C}_S^{(1,0)\ \eta\eta^\prime}$    &$\mathcal{C}_U^{(1,0)\ \eta\eta^\prime} $ &$\mathcal{C}^{(1,0) \ \eta\eta^\prime}_0$             &$\mathcal{C}^{(1,0)\ \eta\eta^\prime}_1$ &$0$ &$\mathcal{C}^{(1,0)\ \eta\eta^\prime}_{35}$
\\
                       &$D^\ast_s\eta^\prime \to D^\ast_s\eta^\prime$      &$0$                             &$2(c_\theta-\sqrt{2}s_\theta)^2$                  &$2(c_\theta-\sqrt{2}s_\theta)^2$                 &$\mathcal{C}^{(1,0)\ \eta^\prime\eta^\prime}_0$ &$\mathcal{C}^{(1,0)\ \eta^\prime\eta^\prime}_1$ &$1$ &$\mathcal{C}^{(1,0)\ \eta^\prime\eta^\prime}_{35}$      
\\ 
\hline 
$(0,\frac12)$  &$D^\ast \pi\to D^\ast \pi$                                         &$-2$                             &$9$     &$-3$  &$M_\pi^2$ &$-M_\pi^2$ &$1$ &$1$
\\
                      &$D^\ast \eta\to D^\ast \eta$                                     &$0$                              &$(c_\theta-\sqrt{2}s_\theta)^2$  &$(c_\theta-\sqrt{2}s_\theta)^2$ &$\mathcal{C}^{(0,\frac12)\ \eta\eta}_0$ &$\mathcal{C}^{(0,\frac12)\ \eta\eta}_1$ &$1$ &$\mathcal{C}^{(0,\frac12)\ \eta\eta}_{35}$ 
\\
                       &$D^\ast_s\bar K\to D^\ast_s\bar K$                        &$-1$                            &$6$& $0$ &$M_{K}^2$ &$-M_{K}^2$ &$1$ &$1$
\\ 
                       &$D^\ast \eta\to D^\ast\pi$                                        &$0$                             &$3(c_\theta-\sqrt{2}s_\theta)$  &$3(c_\theta-\sqrt{2}s_\theta)$  &$0$ &$M_\pi^2(\sqrt{2}s_\theta-c_\theta)$ &$0$ &$c_\theta-\sqrt{2}s_\theta$
\\
                       &$D^\ast_s\bar K\to D^\ast \pi$                                 &$-\frac{\sqrt{6}}{2}$                  &$3\sqrt{6}$  &$0$ &$0$ &$-\frac{\sqrt{6}(M^2_K+M^2_\pi)}{4}$ &$0$ &$\frac{\sqrt{6}}{2}$
\\
                       &$D^\ast_s\bar K\to D^\ast \eta$                              &$-\frac{\sqrt{6}}{2}c_\theta$      &$\sqrt{6}(c_\theta-\sqrt{2}s_\theta)$  
                       &$-2\sqrt{3}(\sqrt{2}c_\theta+s_\theta)$ &$0$ &$\mathcal{C}^{(0,\frac12)\ \bar{K}\eta}_1$ &$0$ &$\mathcal{C}^{(0,\frac12)\ \bar{K}\eta}_{35}$
\\
                       &$D^\ast \eta^\prime \to D^\ast \pi$                           &$0$             &$3(\sqrt{2}c_\theta+s_\theta)$    &$3(\sqrt{2}c_\theta+s_\theta)$  &$0$ &$-M^2_\pi(\sqrt{2}c_\theta+s_\theta)$ &$0$ &$s_\theta+\sqrt{2}c_\theta$
\\
                       &$D^\ast \eta\to D^\ast \eta^{\prime}$                       &$0$             &$\mathcal{C}^{(0,\frac12)\ }_{(c_\theta,s_\theta)}$ &$\mathcal{C}^{(0,\frac12)\ }_{(c_\theta,s_\theta)}$ &$\mathcal{C}^{(0,\frac12)\ \eta\eta^{\prime}}_0$ &$\mathcal{C}^{(0,\frac12)\ \eta\eta^{\prime}}_1$ &$0$ &$\mathcal{C}^{(0,\frac12)\ \eta\eta^{\prime}}_{35}$
\\
                        &$D^\ast_s\bar K\to D^\ast \eta^\prime$                  &$-\frac{\sqrt{6}}{2}s_\theta$   &$\sqrt{6}(\sqrt{2}c_\theta+s_\theta)$  &$2\sqrt{3}(c_\theta-\sqrt{2}s_\theta)$ &$0$ &$\mathcal{C}^{(0,\frac12)\ \bar{K}\eta^{\prime}}_1$ &$0$ &$\mathcal{C}^{(0,\frac12)\ \bar{K}\eta^{\prime}}_{35}$
\\
                         &$D^\ast \eta^\prime \to D^\ast \eta^\prime$           &$0$        &$(\sqrt{2}c_\theta+s_\theta)^2$         &$(\sqrt{2}c_\theta+s_\theta)^2$          &$\mathcal{C}^{(0,\frac12)\ \eta^\prime \eta^\prime}_0$ &$\mathcal{C}^{(0,\frac12)\ \eta^\prime \eta^\prime}_1$                    &$1$           &$\mathcal{C}^{(0,\frac12)\ \eta^\prime \eta^\prime}_{35}$
\\
\noalign{\global\arrayrulewidth=2pt}
\hline
\noalign{\global\arrayrulewidth=0.4pt}
\end{tabular}
}
\end{table*}

The Feynman diagrams for our calculation are displayed in Fig.~\ref{fig:2}. The potentials for the processes of $\mP_1^\ast(p_1)\phi_1(p_2) \rightarrow \mP_2^\ast(p_3)\phi_2(p_4)$ can be expressed as
\begin{align}
V(s,t)=V_{\rm ct}(s,t)+V_{\rm ex}(s,t)\ ,
\end{align}
where the first term corresponds to the contribution from the contact diagrams ($c$) and ($f$), while the second term accounts for that from the exchange diagrams. The explicit expressions of $V_{\rm ct}$ and $V_{\rm ex}$ read
\begin{align}
V_{\rm ct}(s,t)&=-\epsilon_1\cdot\epsilon_3^\ast
\bigg\{\mathcal{C}_\text{LO}\frac{s-u}{4F_0^2}-\frac{1}{F_0^2}\bigg[4h^\ast_0\mathcal{C}_0^{(2)}-
{2}h^\ast_1\mathcal{C}_1^{(2)}+2\mathcal{C}_{24}^{(2)}\mathcal{H}_{24}-
2\mathcal{C}_{35}^{(2)}\mathcal{H}_{35}\bigg]\bigg\}\ ,\label{eq.pot.ct}\\
V_{\rm ex}(s,t)&=\frac{\mathcal{C}_S}{3F_0^2}\bigg[g_0^2\mathcal{G}_{\mathcal{P}}(s,t)-g_1^2\mathcal{G}_{\mathcal{P}^\ast}(s,t)\bigg]+\frac{\mathcal{C}_U}{3F_0^2}\bigg[g_0^2\mathcal{G}_{\mathcal{P}}(u,t)-g_1^2\mathcal{G}_{\mathcal{P}^\ast}(u,t)\bigg]\ ,\label{eq.pot.ex}
\end{align}
where the coefficients $\mathcal{C}_{i}$ ($i=\text{LO},0,1,24,35,S,U$) are compiled in Table~\ref{tab:ciex}. The conservation of charm, strangeness and isospin enforces $\mathcal{C}_S=0$ for the $(S,I)=(-1,0)$, $(-1,1)$, $(2,1/2)$, $(0,3/2)$ and $(1,1)$ channels. Here, $\epsilon_{1}$ and $\epsilon_{3}$ are polarization vectors of the incoming and outgoing vector $\mathcal{P}^\ast$ mesons, respectively. The Mandelstam variables $s,t,u$ are defined by 
\begin{align}
    s &\equiv(p_{1}+p_{2})^2\ ,\quad
    t \equiv(p_{1}-p_{3})^2\ ,\quad
    u \equiv(p_{1}-p_{4})^2\ ,
\end{align}
which fulfill the constraint $s+t+u=m_1^2+m_2^2+m_3^2+m_4^2$, with $p_i^2=m_i^2$ ($i=1,2,3,4$). The functions in the above potentials are given by
\begin{align}
\mathcal{H}_{24}(s,t)&=2h^\ast_2\,p_2\cdot p_4+h^\ast_4\,(p_1\cdot p_2 p_3\cdot p_4+p_1\cdot
p_4 p_2\cdot p_3)\ ,\\
\mathcal{H}_{35}(s,t)&=h^\ast_3\,p_2\cdot p_4+h^\ast_5\,(p_1\cdot p_2 p_3\cdot p_4+p_1\cdot
p_4 p_2\cdot p_3)\ ,\\
\mathcal{G}_{\mathcal{P}}(s,t)&=\frac{p_2\cdot \epsilon_1p_4\cdot\epsilon_3^\ast}{(p_1+p_2)^2-M_{h^2}} \ , \\
\mathcal{G}_{\mathcal{P}^\ast}(s,t)&=\frac{1}{(p_1+p_2)^2-M_{h^{\ast}}^2}
\bigg\{p_1\cdot\epsilon_3^\ast\big[p_2\cdot p_4 p_3\cdot\epsilon_1 -p_2\cdot p_3 p_4\cdot\epsilon_1 \big]+p_2\cdot\epsilon_3^\ast\nonumber\\
&\times\big[p_1\cdot p_3 p_4\cdot\epsilon_1 -p_1\cdot p_4 p_3\cdot\epsilon_1 \big]
+\epsilon_1\cdot\epsilon_3^\ast\big[p_2\cdot p_3 p_1 \cdot p_4 -p_1\cdot p_3 p_2 \cdot p_4 \big]\bigg\}\ ,
\end{align}
with $M_{h}$ ($M_{h^\ast}$) the mass of the intermediate pseudoscalar (vector) charmed meson that appears in the exchange diagrams.

\subsection{Helicity amplitudes}
The helicity formalism was first proposed by Ref.~\cite{Jacob:1959at} for scatterings of particles with spin, facilitating both partial-wave decomposition and the connection to experimental observables~\cite{Martin:1970hmp}. In our case, the helicity amplitudes for the scattering processes of $\mP^\ast_1(p_1)\phi_1(p_2)\to\mP^\ast_2(p_3)\phi_2(p_4)$ are defined by
\begin{align}
V_{\lambda_3\lambda_1}(p_2,p_4;\Sigma)=\epsilon_{\lambda_3}^{\mu\dagger}(p_3)\,V_{\mu\nu}(p_2,p_4;\Sigma)\,\epsilon_{\lambda_1}^\nu(p_1) \ , \label{eq:Lorentzdec}
\end{align}
with the $\lambda_{1,3}\in\{0,\pm1\}$ being the helicities of the initial or final vector mesons. The rank-2 tensor amplitude $V^{\mu\nu}(p_2,p_4;\Sigma)$ can be further decomposed as 
\begin{align}
V^{\mu\nu}&=V_1\,(g^{\mu\nu}-\hat{\Sigma}^\mu\hat{\Sigma}^\nu)+V_2\,\hat{\Sigma}^\mu\hat{\Sigma}^\nu+V_3\,\hat{\Sigma}^\mu p_{\rm 4T}^{\nu}
+V_4\,p_{\rm 2T}^\mu\hat{\Sigma}^\nu+V_5\,p_{\rm 2T}^\mu p_{\rm 4T}^\nu\ ,\label{eq:Lorentz2}
\end{align}
where $\Sigma^\mu=(p_2+p_4)^\mu$ and $\hat{\Sigma}^\mu\equiv {\Sigma^\mu}/{\sqrt{\Sigma^2}}$; the transverse momenta $p_{i{\rm T}}$ are defined by
\begin{align}
 p^\mu_{i{\rm T}}=p_i^\mu-\Sigma^\mu\frac{p_i\cdot\Sigma}{\Sigma^2}\ ,\quad i= 2,4\ .
\end{align}
The coefficients $V_{i}$ ($i=1,\cdots,5$) are Lorentz invariant functions. Their explicit expressions can be derived with the help of the chiral potentials shown in the preceding subsection, which are relegated to Appendix~\ref{Apend:amp}.

For $\mathcal{P}^\ast\phi$ scattering, there are in total $9$ helicity amplitudes. However, only $5$ of them are independent because strong interaction is invariant under parity inversion $\mathbb{P}$ and time reversal $\mathbb{T}$. Specifically, one has 
\begin{align}
\mathbb{P}:\quad  V_{\lambda_3\lambda_1}&=(-1)^{\lambda_3-\lambda_1}V_{-\lambda_3-\lambda_1}\ ,\\
\mathbb{T}:\quad V_{\lambda_3\lambda_1}&=(-1)^{\lambda_3-\lambda_1}V_{\lambda_1\lambda_3}\ ,
\end{align}
which lead to
$V_{++}=V_{--}$, $V_{+-}=V_{-+}$, $V_{+0}=-V_{-0}$ and $V_{0+}=-V_{0-}$, 
where $\pm$ are shorthands for $\pm 1$. The helicity amplitude $V_{00}$ is itself invariant under $\mathbb{P}$ and $\mathbb{T}$ transformations.

In fact, the five independent helicity amplitudes can be expressed in terms of the five covariant amplitudes $V_{i=1,\cdots,5}$, defined in Eq.~\eqref{eq:Lorentz2}. In the center of mass (CM) frame, the relevant four momenta can be expressed as 
\begin{align}
&p_1^\mu=(\omega_{1},0,0,{p}_{\rm cm})^T\ ,\quad p_2^\mu=(\omega_{2},0,0,-{p}_{\rm cm})^T\ , \nonumber \\ 
&p_3^\mu=(\omega_{3},\bar{p}_{\rm cm}\sin\varphi,0,\bar{p}_{\rm cm}\cos\varphi)^T\ , \nonumber \\
\qquad 
&p_4^\mu=(\omega_{4},-\bar{p}_{\rm cm}\sin\varphi,0,-\bar{p}_{\rm cm}\cos\varphi)^T\ ,
\end{align}
with $\omega_{i}$ being the CM energies. In above, $\varphi$ stands for the scattering angle between the incoming and outgoing states. The CM momenta $p_{\rm cm}$ and $\bar{p}_{\rm cm}$ are given by
\begin{align}\label{eq.pcms}
p_{\rm cm} = \frac{\lambda^{\frac{1}{2}}(s,m_1^2,m_2^2)}{2 \sqrt{s}} \ , \quad 
\bar{p}_{\rm cm} = \frac{\lambda^{\frac{1}{2}}(s,m_3^2,m_4^2)}{2 \sqrt{s}}\ ,
\end{align}
with $\lambda(a,b,c)\equiv a^2+b^2+c^2-2ab-2bc-2ca$ the K\"all\'en function. In the CM frame, the polarization vectors of the vector charmed mesons can be expressed as
\begin{align}
&\epsilon_{\pm1}^\mu(p_1)=(0,\frac{\mp1}{\sqrt{2}},\frac{-i}{\sqrt{2}},0)^T ,\quad
\epsilon_0^\mu(p_1)=(\frac{p_{\rm cm}}{m_1},0,0,\frac{\omega_{1}}{m_1})^T , \notag\\
&\epsilon_{\pm1}^\mu(p_3)=(0,\frac{\mp\cos\varphi}{\sqrt{2}},\frac{-i}{\sqrt{2}},\frac{\pm\sin\varphi}{\sqrt{2}})^T\ ,\quad
\epsilon_0^\mu(p_3)=(\frac{\bar{p}_{\rm cm}}{m_3},\frac{\omega_3\sin\varphi}{m_3},0,\frac{\omega_{3}\cos\varphi}{m_3})^T\ .
\end{align}
Substituting the momenta and polarization vectors into Eqs.~\eqref{eq:Lorentzdec} and \eqref{eq:Lorentz2}, one obtains the relations between the helicity amplitudes and the covariant amplitudes:
\begin{align}
&V_{++}=V_{--}=-\frac{1}{2}\left[(z_s+1){V_1}+p_{\rm cm}\bar{p}_{\rm cm}(1-z_s^2)V_5\right]\ ,\notag\\
&V_{+-}=V_{-+}=\frac{1}{2}\left[(z_s-1)V_1+p_{\rm cm}\bar{p}_{\rm cm}(1-z_s^2)V_5\right]\ ,\notag\\
&V_{+0}=-V_{-0}=-{\frac{{(1-z_s^2)^{\frac{1}{2}}}}{\sqrt{2}m_1}}\left[\omega_1V_1 - p_{\rm cm}^2V_4-z_s\omega_1p_{\rm cm}\bar{p}_{\rm cm}V_5\right]\ ,\notag\\
&V_{0+}=-V_{0-}={\frac{(1-z_s^2)^{\frac{1}{2}}}{\sqrt{2}m_3}}\left[\omega_3V_1 -\bar{p}_{\rm cm}^2V_3-z_s\omega_3p_{\rm cm}\bar{p}_{\rm cm}V_5\right]\ ,\notag\\
&V_{00}=\frac{1}{m_1m_3}\big[z_sp_{\rm cm}^2\omega_3V_4 + z_s\omega_1(\bar{p}_{\rm cm}^2V_3-\omega_3V_1)+p_{\rm cm}\bar{p}_{\rm cm}(V_2+z_s^2\omega_1\omega_3V_5)\big]\ ,
\label{eq:helicity-Lorentz}
\end{align}
with $z_s\equiv\cos\varphi$. Consequently, chiral expressions of the above helicity amplitudes follow from those of the invariant amplitudes $V_{i}$ ($i=1,\cdots,5$); see Eqs.~\eqref{eq.inv.amp.V1}-\eqref{eq.inv.amp.V5}.

\subsection{Partial wave amplitudes}
The partial wave decomposition of the helicity amplitudes has the following form
\begin{align}
V_{{\lambda}_3\lambda_1}(s,t)=\sum_{J=\max\{|{\lambda}_3|,|\lambda_1|\}}^{\infty}\hspace{-0.75cm}(2J+1)V^J_{{\lambda_3}\lambda_1}(s)d^{J}_{\lambda_1{\lambda}_3}(z_s)\ ,
\end{align}
where $d^{J}_{\lambda_1{\lambda_3}}(z_s)$ is the standard Wigner $d$ function and the partial waves are given by
\begin{align}\label{eq.helicity.PW}
V^J_{{\lambda}_3\lambda_1}(s)=\frac{1}{2}\int_{-1}^{1} \rd z_{s} V_{{\lambda}_3\lambda_1}(s,t(s,z_s))\,d^{J}_{\lambda_1{\lambda_3}}(z_s)\ .
\end{align}
The Mandelstam variable $t$ is related to $z_s$ through
\begin{align}
t(s,z_s) =m_1^2+m_3^2+\frac{(s+m_1^2-m_2^2)(s+m_3^2-m_4^2)}{2s}
+2p_{\rm cm}\bar{p}_{\rm cm}z_s\ ,
\end{align}
with $p_{\rm cm}$ and $\bar{p}_{\rm cm}$ given by Eq.~\eqref{eq.pcms}.
However, helicity states are not eigenstates of parity reflection. As a result, the partial wave helicity amplitudes do not possess definite parity quantum numbers. 

To obtain partial wave amplitudes with definte $J^{P}$, one can employ the so-called $JLS$ basis. This can be achieved via the following transformation,
\begin{align}
{V}_{JLS} = \sum_{\lambda_1 {\lambda_3}}\mathcal{U}_{\lambda_3 0}^{JLS} V_{\lambda_3{\lambda_1}}^{J}[\mathcal{U}_{\lambda_1 0 }^{JLS}]^\dagger\ ,
\end{align}
with $S=S_1=S_3=1$ and $J=L+S$. The parity of the above amplitude is $P=(-1)^{L}$. Note that since the total spin of the $\mathcal{P}^\ast\phi$ system is always $S=1$, the abbreviation $ V_{JL}(s)$ will be used hereafter. The components of the matrix $\mathcal{U}^{JLS}$ are given by 
\begin{align} 
\mathcal{U}_{\lambda_{i} 0}^{JLS} = \bigg(\frac{2L+1}{2S+1}\bigg)^{\frac{1}{2}}
\langle L0S\lambda|J\lambda\rangle
\langle S_{i} \lambda_{i} 00 | S  \lambda \rangle \ ,
\end{align}
with $i=1$ or $3$, whose values are determined by the pertinent Clebsch-Gordan coefficients. 

For $S$ wave with $L=0$ and $J=S=1$, the transformation matrix reads
\begin{align}
\mathcal{U}^{101}   =(\mathcal{U}_{+ 0}^{101},\mathcal{U}_{0 0}^{101},\mathcal{U}_{- 0}^{101}) =(\frac{1}{\sqrt{3}},\frac{1}{\sqrt{3}},\frac{1}{\sqrt{3}})\ .
\end{align}
Consequently, the $S$-wave amplitude with definite quantum numbers $J^P=1^+$ takes the general form
\begin{align}
{V}_{J=1,L=0}(s)=\frac{1}{3}\,\big[2V_{++}^{1}(s)+2V_{+0}^{1}(s)+2V_{+-}^{1}(s)
+2V_{0+}^{1}(s)+V_{00}^{1}(s)\big]  \ ,
\end{align}
where $V_{\lambda_3\lambda_1}^J$ are the partial-wave helicity amplitudes defined in Eq.~\eqref{eq.helicity.PW}. For elastic scattering processes at threshold, the CM momenta satisfy $p_{\rm cm}=\bar{p}_{\rm cm}=0$, the corresponding $S$-wave amplitudes become independent of $z_s$ and the potentials $V_{2,3,4,5}(s,t)$. This yields the simplified expression $V_{J=1,L=0}(s_{\rm th})=-V_1(s_{\rm th},t(s_{\rm th}))$.

Likewise for the $P$ wave with $L=1$, the total angular momenta can be $J=0,1,2$, and the corresponding transformation matrices can be derived as
\begin{align}
 \mathcal{U}^{011}   &=(\mathcal{U}_{+ 0}^{011},\mathcal{U}_{0 0}^{011},\mathcal{U}_{- 0}^{011}) =(0,-1,0)\ ,\\
 \mathcal{U}^{111}   &=(\mathcal{U}_{+ 0}^{111},\mathcal{U}_{0 0}^{111},\mathcal{U}_{- 0}^{111}) =(-{1}/{\sqrt{2}},0,{1}/{\sqrt{2}})\ ,\\
 \mathcal{U}^{211}   &=(\mathcal{U}_{+ 0}^{211},\mathcal{U}_{0 0}^{211},\mathcal{U}_{- 0}^{211}) =(\sqrt{{3}/{10}},\sqrt{{2}/{5}},\sqrt{{3}/{10}})\ .
\end{align}
Then, the $P$-wave amplitudes in $JLS$ basis are given by 
\begin{align}
V_{J=0,L=1}(s)&=V_{00}^0(s)  \ ,\\
V_{J=1,L=1}(s)&=V_{++}^1(s)-V_{+-}^1(s) \ ,\\ 
V_{J=2,L=1}(s)&=\frac15\big[2V_{00}^2(s)+3\big(V_{++}^2(s)+V_{+-}^2(s)\big)
+2\sqrt{3}\big(V_{0+}^2(s)+V_{+0}^2(s)\big)\big] \ , 
\end{align}
corresponding to $J^P=0^-,1^-,2^-$, respectively.

\subsection{Unitarization}

The partial wave amplitudes derived from chiral effective field theory respect only perturbative unitary. In order to scrutinize dynamically generated pole structures, the unitarity must be restored. A popular manner is to conduct unitarization using the BSE under on-shell approximation~\cite{Oller:2000fj,Oller:1997ti}. Specifically, the formulation of the BSE approach reads
\begin{align}
\mathcal{T}_{JL}^{(S,I)}(s)=\mathcal{V}_{JL}^{(S,I)}(s)\cdot\left[1-\mathcal{G}(s)\cdot\mathcal{V}_{JL}^{(S,I)}(s)\right]^{-1}\ .\label{eq.BSE.matrix}
\end{align}
Here $\mathcal{V}_{JL}^{(S,I)}(s)$ and $\mathcal{G}(s)$ are expressed in the matrix notation and take the form
\begin{align}
\mathcal{V}_{JL}^{(S,I)}=
\left(
\begin{array}{cccc}
\left[{V}_{JL}^{(S,I)}\right]_{1\to 1}    &\cdots     &\left[{V}_{JL}^{(S,I)}\right]_{1\to j}     &\cdots  \\ 
\vdots                                               &\ddots     &\vdots                                               &\ddots \\
\left[{V}_{JL}^{(S,I)}\right]_{i\to 1}     &\cdots     &\left[{V}_{JL}^{(S,I)}\right]_{i\to j}      &\cdots  \\
\vdots                                               &\ddots     &\vdots                                               &\ddots \\
\end{array}
\right)\ ,\quad
\mathcal{G}(s)={\rm diag}\{g_1(s),\cdots,g_{i}(s),\cdots\}\ ,\label{eq.G.matrix}
\end{align}
where $i$ and $j$ are channel indices, and ${V}_{JL}^{(S,I)}$ represents the $L$-wave projection of the $\mathcal{O}(p^2)$ scattering potential with strangeness $S$ and isospin $I$. Taking the processes with $(S,I)=(1,1)$ for example, there are two channels: $1=D_s^\ast\pi$ and $2=D^\ast K$. The $L$-wave potential matrix reads
\begin{align}
  \mathcal{V}_{JL}^{(1,1)}=
\left(
\begin{array}{cc}
\left[{V}_{JL}^{(1,1)}\right]_{D_s^\ast\pi\to D_s^\ast\pi}       &\left[{V}_{JL}^{(1,1)}\right]_{D_s^\ast\pi\to D^\ast K}       \\ 
\left[{V}_{JL}^{(1,1)}\right]_{D_s^\ast\pi\to D^\ast K}        &\left[{V}_{JL}^{(1,1)}\right]_{D^\ast K\to D^\ast K}       \\
\end{array}
\right)\ .  \notag
\end{align}
Note that, for single channels with $(S,I)=(-1,0)$, $(-1,1)$, $(2,\frac{1}{2})$ and $(0,\frac{3}{2})$ (see Table~\ref{tab:ciex}),  the above matrix equation~\eqref{eq.BSE.matrix} reduces to simple algebraic equation. 

The two-point one-loop function $g_{i}(s)$ in Eq.~\eqref{eq.G.matrix} is defined by~\cite{Oller:1998zr}
\begin{align}
g_i(s) ={i}\int\frac{{\rm d}^4 k}{(2\pi)^4}\frac{1}{[k^2-M_{\mathcal{P}_i^\ast}^2+i \varepsilon][(k+P)^2-M_{\phi_i}^2+i\varepsilon]}\ ,\notag
\end{align}
with $s\equiv P^2$ and $\varepsilon$ an infinitesimal positive number. $g_i(s)$ collects unitary cuts generated by the two-particle intermediate states, i.e., $\mathcal{P}^\ast_i\phi_i$ states, in the $i$-th channel. It can be calculated in dimensional regularization by replacing the ultraviolet divergence with a constant, leading to
\begin{align}
g_i(s)&=\frac{1}{16\pi^2}\bigg\{{a}_i(\mu)+\ln\frac{M_{\mP^{\ast}_i}^2}{\mu^2}+\frac{s-M_{\mP^{\ast}_i}^2+M_{\phi_i}^2}{2s}\ln\frac{M_{\phi_i}^2}{M_{\mP^{\ast}_i}^2}\nonumber\\
&+\frac{\sigma_i(s)}{2s}\big[\ln(s-M_{\phi_i}^2+M_{\mP^{\ast}_i}^2+\sigma_i(s))-\ln(-s+M_{\phi_i}^2-M_{\mP^{\ast}_i}^2+\sigma_i(s))\nonumber\\
&\hspace{1cm}+\ln(s+M_{\phi_i}^2-M_{\mP^{\ast}_i}^2+\sigma_i(s))-\ln(-s-M_{\phi_i}^2+M_{\mP^{\ast}_i}^2+\sigma_i(s))\big]\bigg\}\ ,
\label{eq:g}
\end{align}
with $\sigma_i(s)\equiv\lambda^{1/2}(s,M_{\mP^{\ast}_i}^2,M_{\phi_i}^2)$. Here, the constant $a_{i}(\mu)$ depends on the renormalization scale $\mu$, which is usually set equal to the chiral symmetry breaking scale: $\mu=4\pi F_\pi\sim 1$~GeV. Nevertheless, the $g_i(s)$ is $\mu$-independent, since the $\mu$-dependences of the first two terms in the first line of Eq.~\eqref{eq:g} cancel with each other.

\section{Numerical setup and scattering lengths}
\label{sec:numerical}

\subsection{Parameters \label{subs:CC}}

The physical masses used in this work are tabulated in Table.~\ref{cof:mass}. We work in the exact isospin limit, which implies that the states in an isospin multiplet are mass-degenerate. Therefore, the pion masses are uniformly set to $M_\pi=138.04$~MeV, the average of the physical masses of charged and neutral pions in the pion triplet. The mass values of $K$, $\bar{K}$, $D$ and $D^\ast$ are obtained in the same manner.  Furthermore, following Ref.~\cite{Liu:2012zya}, we take  $F_0=F_\pi$ with the pion decay constant $F_\pi=92.2~\rm{MeV}$~\cite{ParticleDataGroup:2026aaa}.

\begin{table}[htbp]
\caption{Physical masses used in this work. } \label{cof:mass}
\centering
\begin{tabular}{c|c c c c c c}
\noalign{\global\arrayrulewidth=2pt}
\hline
\noalign{\global\arrayrulewidth=0.4pt}
$\phi$                    &$M_{\pi}$      &$M_{K}$       &$M_{\eta}$            &$M_{\eta^{\prime}}$    \\
Mass~[MeV]          &$138.04$      &$495.16$    &$547.86$              &$957.78$ \\
\hline
$\mP^{\ast}/\mP$  &$M_{D^{\ast}}$      &$M_{D_s^{\ast}}$ &$M_{D}$       &$M_{D_s}$   \\
Mass~[MeV]          &$2008.55$            &$2112.2$               &$1867.24$    &$1968.34$  \\                            
\noalign{\global\arrayrulewidth=2pt}
\hline
\noalign{\global\arrayrulewidth=0.4pt}
\end{tabular}
\end{table}

The LECs $h_i^\ast$ ($i=0,\cdots,5$) involved in the NLO contact $\mathcal{P}^\ast\phi$ scattering potentials have not yet been determined so far, due to the lack of experimental data and insufficient lattice QCD data. The value of $h^{\ast}_{1}$ can be fixed by the physical mass difference between $D^{\ast}$ and $D^{\ast}_{s}$. Specifically, the NLO $D^{\ast}$ and $D^{\ast}_{s}$ masses read
\begin{align}
&M^{2}_{D^\ast}=\overline{M}^{2}_{\mP^\ast} + 2(h^\ast_{0}+h^\ast_{1}) M^{2}_{\pi} + 4h^\ast_{0}M^{2}_{K}  \ , \label{eq.mass.Dst}\\
&M^{2}_{D^\ast_{s}}=\overline{M}^{2}_{\mP^\ast} + 2(h^\ast_{0}-h^\ast_{1}) M^{2}_{\pi} + 4(h^\ast_{0}+h^\ast_{1})M^{2}_{K}  \ ,\label{eq.mass.Dsst}
\end{align}
which yield
\begin{align}
h^{\ast}_1=\frac{M^2_{D^\ast_s}-M^2_{D^\ast}}{4(M_K^2-M_\pi^2)} = 0.47 \ .\label{eq.lec.h1ast}
\end{align}
For the remaining $h^\ast_{i}$, the HQSS leads to 
\begin{align}
h^{\ast}_{i}=h_{i},\,\quad i =0, 2,3,4,5\ ,\label{eq.HQSS.relations}
\end{align}
where $h_i$ are the NLO LECs appearing in $\mathcal{P}\phi$ scattering. With HQSS, the scatterings of pNGBs off vector $\mathcal{P}^\ast$ mesons and off pseudoscalar $\mathcal{P}$ mesons are related to each other. Fortunately, the HQSS appears to be reasonably well satisfied~\cite{Du:2017zvv}, even at a large unphysical pion mass of $\sim 391$~MeV~\cite{Lang:2022elg}, ensuring that the relations in Eq.~\eqref{eq.HQSS.relations} hold well.\footnote{In fact, $h_1$ can be alternatively estimated by the physical mass difference between $D$ and $D_{s}$, in the same way as Eq.~\eqref{eq.lec.h1ast} for $h_1^\ast$, and the resulting value is $h_1=0.43$. Therefore, the breaking effect of HQSS is roughly $\delta=|h_1^\ast-h_1|/|h_1^\ast|\sim 8.5\%$.}

\begin{table}[hbp]
\caption{Values of the NLO LECs appearing in the $\mathcal{P}^\ast\phi$ contact interaction. } \label{value:LECs}
\centering
\begin{tabular}{c|cccccc}
\noalign{\global\arrayrulewidth=2pt}
\hline
\noalign{\global\arrayrulewidth=0.4pt}
LECs      & 
$h_0^\ast$ &
$h_1^\ast$ &
$h_2^\ast$ &
$h_3^\ast$ &
$h_4^\ast$ &
$h_5^\ast$  \cr
\hline
Value 
&$0.033$ 
&$0.47$  
&$0.08^{+0.31}_{-0.34}$
&$3.79^{+0.41}_{-0.41}$
&$-0.06^{+0.08}_{-0.07}$ 
&$-0.48^{+0.05}_{-0.05}$ \cr
\noalign{\global\arrayrulewidth=2pt}
\hline
\noalign{\global\arrayrulewidth=0.4pt}
\end{tabular}
\end{table}

For the $\mathcal{P}\phi$ interactions, the NLO LECs $h_i$ were first estimated in Ref.~\cite{Guo:2009ct} by means of $1/N_C$ arguments and naturalness assumptions. Later, they were better determined by fitting to lattice QCD data using SU(3) chiral potentials of NLO~\cite{Liu:2012zya,Guo:2018tjx} and next-to-next-to leading order (NNLO)~\cite{Yao:2015qia,Du:2017ttu}. Further, stringent positivity bounds on those LECs were derived based on the general principles of $S$-matrix theory including analyticity, crossing symmetry and unitarity~\cite{Du:2016tgp}. The SU(3) ChPT study of low-energy $\mathcal{P}\phi$ scatterings was extended to the U(3) framework in Ref.~\cite{Guo:2015dha}. Here, since we also work in U(3) ChPT, we take the values of $h_i$ from Ref.~\cite{Guo:2015dha} (Fit-6C of Table III therein) to estimate those of $h_i^\ast$ via Eq.~\eqref{eq.HQSS.relations}. The NLO $\mathcal{P}^\ast\phi$ LEC results are summarized in Table~\ref{value:LECs}, where the quoted errors are statistical. Note that, in what follows, we employ a common subtraction constant $a(\mu)$ in $g_i(s)$ of Eq.~\eqref{eq:g} for all the scattering channels, whose value is set to $a(\mu=1~\text{GeV})=-1.88(6)$, taken from Fit-6C of Table III of Ref.~\cite{Guo:2015dha}.

As for the coupling constant $g_{0}$ in the meson-exchange diagrams, its value can be extracted from the decay width of $D^{\ast +}(2010) \to D^{0} \pi^{+}$. At LO, the decay width is
\begin{align}
\Gamma(D^{\ast +}(2010) \to D^{0} \pi^{+}) = \frac{g_{0}^2}{12 \pi F_{\pi}^2} \frac{ |\bm{q}_{\pi}|^3 }{M_{D^{\ast +}}^2} \ ,
\end{align}
with $\bm{q}_{\pi}$ the pion three-momentum in the rest frame of the mother particle $D^{\ast+}$. Together with the total decay width $\Gamma = 83.4 \pm 1.8~\rm{keV}$ and the branching fraction ${\text Br}(D^{\ast+}(2010) \to D^{0} \pi^{+})=(67.7\pm 0.5)~\%$~\cite{ParticleDataGroup:2026aaa}, the $g_0$ is estimated to be $g_0=1095.0 \pm 15.8$~\rm{MeV}. For $g_1$, the coupling of $\mathcal{P}^\ast \mathcal{P}^\ast\phi$ interaction, we impose the HQSS relation $g_{1} M_{D^{\ast}} = g_{0}$~\cite{Jiang:2019hgs} to determine its value.

\subsection{Prediction of $S$-wave scattering lengths}

Once the values of all the involved parameters are pinned down, one can make predictions for the scattering lengths. For the elastic process of $\mP^\ast_i\phi_i \rightarrow \mP^\ast_i \phi_i$, the corresponding scattering length is related to its unitarized scattering amplitude through
\begin{align}
a^{(S,I)}_{JL}(\mP^\ast_i\phi_i\to \mP^\ast_i\phi_i)
=-\frac{1}{8\pi\sqrt{s_{\rm th}^i}} \lim_{s\to s_{\rm th}^i}\bigg[\frac{\mathcal{T}^{(S,I)}_{JL, \mP^\ast_i\phi_i \to \mP^\ast_i\phi_i}(s)}{p_{i,{\rm cm}}^{2L}} \bigg]\ , \label{eq.sl.for}
\end{align}
where $s_{\rm th}^i=(M_{\mP^\ast_i} +M_{\phi_i})^2$ denotes the threshold in the $i$-th channel. In above, the sign convention is chosen such that a negative (positive) scattering length corresponds to an attractive (repulsive) interaction. 

\begin{table}[htbp] 
\caption{$S$-wave scattering lengths for the elastic scattering channels in units of fm. The results determined in HMChPT~\cite{Liu:2011mi} are also shown for easy comparison.}
\label{Tab:sl}
\renewcommand{\arraystretch}{1.0}
\begin{tabular}{ l | c| c  c | c}
\noalign{\global\arrayrulewidth=2pt}
\hline
\noalign{\global\arrayrulewidth=0.4pt}
$(S,I)$ & Process &CT     & CT + Ex.       &  HMChPT~\cite{Liu:2011mi}
\\
\hline
$(-1,0)$ & $D^{\ast}\bar{K}\to D^{\ast}\bar{K}$ &$\;\;\;0.41(2)$ &$\;\;\;0.41(2)$
& $0.29+i\,5.2\times 10^{-6}$\\
$(-1,1)$ & $D^{\ast}\bar{K}\to D^{\ast}\bar{K}$ &$-0.18(1)$   &$-0.18(1)$ & $-0.19-i\,1.7\times 10^{-6}$
\\
$(2,\frac12) $& ${D^{\ast}_s K}\to{D^{\ast}_s K} $ &$-0.17(1)$ &$-0.17(1)$ & $-0.14$
\\
${(0,\frac32)}$& ${D^{\ast} \pi }\to{D^{\ast} \pi } $ &$-0.10(1)$ &$-0.10(1)$ & $-0.13-i\,3.6\times 10^{-4}$
\\
\hline
${(1,1)}$ & ${D^{\ast}_s \pi }\to {D^{\ast}_s \pi }$ &$-0.001(1)$ &$-0.004(1)$ & $-0.039$
\\ 
& ${D^{\ast}K}\to {D^{\ast}K}$ &$0.01(1)+i\,0.07(1)$ &$0.01(1)+i\,0.05(1)$ & 
$-0.022+i\,0.10$
\\ \hline
${(1,0)} $ & ${D^{\ast} K}\to {D^{\ast} K} $ &$-0.86^{+0.09}_{-0.14}$ &$-0.85^{+0.09}_{-0.14}$  & $0.76-i\,5.2\times 10^{-6}$
\\
& ${D^{\ast}_s \eta }\to {D^{\ast}_s \eta }$ &$-0.22(1)+i\,0.04(1)$ &$-0.22(1)+i\,0.04(1)$ & $0.18+i\,0.19$
\\
& ${D^{\ast}_s \eta^{\prime}}\to {D^{\ast}_s \eta^{\prime}}$ &$-0.24(2)+i\,0.03(1)$ &$-0.26(2)+i\,0.04(1)$ & -
\\ \hline
${(0,\frac12)}$ & ${D^{\ast} \pi}\to{D^{\ast} \pi}$ &$0.35(1)$ &$0.36^{+0.01}_{-0.02}$ & $0.27-i\,3.6\times10^{-4}$
\\
& ${D^{\ast} \eta}\to {D^{\ast} \eta}$ &$-0.03(2)+i\,0.04(1)$ &$-0.05(2)+i\,0.04(1)$ & $0.051+i\,0.094$
\\
& ${D^{\ast}_s\bar{K}}\to {D^{\ast}_s\bar{K}}$ &$-0.12(3)+i\,0.18^{+0.04}_{-0.03}$ &$-0.14^{+0.03}_{-0.02}+i\,0.16^{+0.04}_{-0.03}$ & $0.35+i\,0.27$
\\
& ${D^{\ast} \eta^\prime}\to {D^{\ast} \eta^\prime}$ &$-0.16(2)+i\,0.01(1)$ &$-0.16(2)+i\,0.01(1)$ & -
\\    
\noalign{\global\arrayrulewidth=2pt}
\hline
\noalign{\global\arrayrulewidth=0.4pt}
\end{tabular}
\end{table}

Our predictions for the $S$-wave scattering lengths are given in Table~\ref{Tab:sl}, where the errors are obtained by varying the LECs in their 1-$\sigma$ uncertainties. The most attractive channel is $D^\ast K\to D^\ast K$ with $(S,I)=(1,0)$, for which the scattering length is the most negative, i.e., $a_{D^{\ast} K}^{(1,0)}=-0.86^{+0.09}_{-0.14}$. As will be discussed in section~\ref{sec.poles}, this channel is sufficiently attractive to form a bound state, which can be identified as $D_{s1}(2460)$. For each scattering length in the table, the contribution from the contact term is shown in the second column, while the sum of the contact term and meson-exchange contributions is displayed in the third column. It can be found that the meson-exchange diagrams contribute negligibly in all scattering channels.

For comparison, we also show the $S$-wave $\mathcal{P}^\ast\phi$ scattering lengths derived within heavy meson ChPT (HMChPT) at NNLO~\cite{Liu:2011mi}.\footnote{Such a comparison is very rough and must be taken with caution. Our results are obtained in a non-perturbative way in the sense that an infinite number of the $s$-channel bubble loops are considered via the BSE unitarization, while those in Ref.~\cite{Liu:2011mi} come from a perturbative HMChPT calculation truncated at NNLO.} Therein, the LECs are either estimated via resonance saturation model or assumed to be zero under the naturalness ansatz. Our predictions for the single channels, $I=0$ $D^\ast\bar{K}$, ${I=1}$ $D^\ast\bar{K}$, ${I=1/2}$ $D_s^\ast K$ and ${I=3/2}$ $D^\ast\pi$, are more or less comparable to the HMChPT ones. However, the scattering lengths in the $(S,I)=(1,0)$ coupled channels exhibit large deviations, which might be caused by the poor chiral convergency due to large kaon and eta masses. That is, higher order terms beyond NNLO in chiral perturbation expansion may contribute sizably. A resummation of the higher order terms is required in light of the existence of $D_{s1}(2460)$ state near threshold. Nevertheless, all those terms are dropped in the perturbative HMChPT calculation~\cite{Liu:2011mi}. In our case, their contributions have been partially taken into account by means of the on-shell BSE unitarization approach.  

Very recently, the two-particle momentum correlation functions between vector charm meson and light-pseudoscalar bosons have been measured by ALICE collaboration~\cite{ALICE:2024bhk}. This has enabled the first experimental extraction of $D^\ast\pi$ and $D^\ast K$ scattering lengths. The extracted residual strong interaction between $D^\ast$ and $\pi$ is found to be negligible, yielding $a_{D^\ast\pi}=0.05\pm0.04({\rm stat})\pm0.02({\rm syst})$ for $(S,I)=(0,3/2)$ and $a_{D^\ast\pi}=-0.03\pm0.05({\rm stat})\pm0.02({\rm syst})$ for $(S,I)=(0,1/2)$. A tension emerges between the experimental results and our NLO predictions, similar to that observed in the $D \pi$ scattering lengths~\cite{Yan:2024yuq}. 

\begin{sidewaystable*}[htbp] 
\centering
\caption{$P$-wave scattering lengths at physical masses for the elastic scattering channels in units of 100 fm$^{3}$. The $P$ waves are distinguished by the quantum numbers $J^P$ together with the standard spectroscopic notation $^{2S+1}L_J$. 
}
\label{tab:sl.p}
\renewcommand{\arraystretch}{1.2}
\renewcommand{\tabcolsep}{0.001pc}
{
\begin{tabular}{ l l | c  c|  c  c | c  c}
\noalign{\global\arrayrulewidth=2pt}
\hline
\noalign{\global\arrayrulewidth=0.4pt}
\multirow{2}*{$(S,I)$} & \multirow{2}*{Process} &\multicolumn{2}{c|}{$0^-$ [$^3P_0$]}     &\multicolumn{2}{c|}{$1^-$ [$^3P_1$]}     &\multicolumn{2}{c}{$2^-$ [$^3P_2$]}
\\
\cline{3-8}
    &         & CT     & CT + Ex.     & CT     & CT+Ex.      &      CT       & CT+Ex. 
\\
\hline
${(-1,0)}$ & ${D^{\ast}\bar{K}}\to{D^{\ast}\bar{K}}$ &$-4.9_{-1.0}^{+1.1}$ &$-4.7_{-1.0}^{+1.1}$   &$-4.6_{-1.0}^{+1.1}$ &$-4.4_{-1.0}^{+1.1}$    &$-4.6_{-1.0}^{+1.1}$    &$-5.8_{-1.0}^{+1.1}$
\\
${(-1,1)}$ & ${D^{\ast}\bar{K}}\to {D^{\ast}\bar{K}}$ &$4.4_{-0.7}^{+0.8}$  &$4.2_{-0.7}^{+0.8}$  &$4.1(7)$ &$3.9(7)$  &$4.1(7)$    &$5.3(7)$
\\
${(2,1/2)} $ & ${D^{\ast}_s K}\to {D^{\ast}_s K} $ &$4.2(7)$  &$4.0(7)$ &$3.9(7)$ &$3.6(7)$  &$3.9(7)$  &$6.1(7)$
\\
${(0,3/2)}$ & ${D^{\ast} \pi }\to {D^{\ast} \pi }$ &$4.2_{-0.7}^{+0.8}$  &$-101.8_{-0.7}^{+0.8}$ &$4.0(7)$  &$106.1(7)$  &$4.0(7)$     &$-93.0(7)$
\\
\hline
${(1,1)}$ & ${D^{\ast}_s \pi }\to {D^{\ast}_s \pi }$ &$-0.4_{-0.7}^{+0.8}$ &$-0.5(7)$ &$-0.4_{-0.7}^{+0.8}$  &$-32.9_{-0.8}^{+0.9}$  &$-0.4_{-0.7}^{+0.8}$   &$-6.4_{-0.6}^{+0.7}$
\\ 
&${D^{\ast}K}\to {D^{\ast}K}$ &$0.6(7)+i\,1.0(2)$   &$0.5(7)+i\,0.9(2)$ &$0.5(7)+i\,0.9(2)$   & $0.4(7)+i\,0.7(2)$   & $0.5(7)+i\,0.9(2)$  &$1.4_{-0.6}^{+0.7}+i\,2.0(3)$
\\ \hline
{${(1,0)} $} & ${D^{\ast} K}\to{D^{\ast} K} $ &$13.3(1.6)$  &$5.4_{-1.2}^{+1.3}$ &$14.1^{+1.6}_{-1.5}$ &$10.7(1.9)$  &$13.9^{+1.6}_{-1.5}$ &$13.2(1.4)$
\\
 & ${D^{\ast}_s \eta }\to{D^{\ast}_s \eta }$ &$3.5^{+0.7}_{-0.6}+i\,0.2(1)$   &$2.6_{-0.6}^{+0.7}+i\,0.04(0)$  &$3.4^{+0.7}_{-0.6}+i\,0.4(1)$     &$2.8_{-0.6}^{+0.7}+i\,0.02(0)$     &$3.4^{+0.7}_{-0.6}+i\,0.4(1)$     &$4.3_{-0.6}^{+0.7}+i\,0.2(1)$
\\
& ${D^{\ast}_s \eta^{\prime}}\to{D^{\ast}_s \eta^{\prime}}$ &$-5.5^{+1.7}_{-1.5}+i\,7.9_{-2.1}^{+2.4}$  &$-2.9_{-2.2}^{+2.6}+i\,12.6_{-3.2}^{+2.8}$   &$-4.7_{-1.6}^{+1.7}+i\,5.3_{-1.5}^{+1.7}$  & $-4.8_{-1.5}^{+1.7}+i\,7.1_{-2.0}^{+2.3}$    &$-4.9_{-1.6}^{+1.8}+i\,5.2_{-1.5}^{+1.8}$      &$-6.4_{-1.9}^{+2.0}+i\,6.6_{-1.9}^{+2.3}$
\\ \hline
${(0,1/2)}$ & ${D^{\ast} \pi}\to {D^{\ast} \pi}$ &$7.9(1.0)$  &$57.9_{-1.4}^{+1.6}$    &$8.4_{-1.0}^{+1.1}$  &$-16.1_{-3.9}^{+4.5}$  & $8.2(1.0)$   &$55.4_{-0.8}^{+0.9}$
\\
& ${D^{\ast} \eta}\to {D^{\ast} \eta}$ &$0.6_{-1.1}^{+0.8}+i\,8.2_{-1.6}^{+1.7}$ &$3.4_{-0.3}^{+0.1}+i\,5.0_{-1.4}^{+1.7}$    &$-0.2^{+0.9}_{-1.1}+i\,7.5_{-1.5}^{+1.6}$   &  $2.4_{-0.5}^{+0.2}+i\,5.9_{-1.5}^{+1.7}$  &$-0.4_{-1.2}^{+0.9}+i\,7.6_{-1.5}^{+1.6}$       &$-0.6_{-1.5}^{+1.2}+i\,10.8(1.9)$
\\
& ${D^{\ast}_s\bar{K}}\to {D^{\ast}_s\bar{K}}$ &$1.5(7)+i\,5.1_{-0.9}^{+1.0}$  &$3.1(3)+i\,3.0_{-1.0}^{+1.1}$     &$0.9_{-0.8}^{+0.9}+i\,5.1(9)$  & $2.4_{-0.5}^{+0.4}+i\,4.1(1.0)$    &$0.8_{-0.8}^{+0.9}+i\,5.1(9)$      &$1.3(8)+i\,5.4(9)$
\\
& ${D^{\ast} \eta^\prime}\to{D^{\ast} \eta^\prime}$ &$-4.2_{-1.2}^{+1.3}+i\,2.6_{-0.5}^{+0.6}$    &$-4.6_{-1.3}^{+1.4}+i\,3.8_{-0.8}^{+0.9}$   &$-3.5_{-1.1}^{+1.2}+i\,2.0_{-0.4}^{+0.5}$  & $-3.9_{-1.2}^{+1.3}+i\,2.5_{-0.5}^{+0.6}$     &$-3.6_{-1.1}^{+1.3}+i\,1.9_{-0.4}^{+0.5}$      &$-4.5_{-1.2}^{+1.4}+i\,2.3_{-0.5}^{+0.6}$
\\    
\noalign{\global\arrayrulewidth=2pt}
\hline
\noalign{\global\arrayrulewidth=0.4pt}
\end{tabular}
}
\end{sidewaystable*}

The $P$-wave scattering lengths are also calculated using Eq.~\eqref{eq.sl.for} and the results are summarized in Table~\ref{tab:sl.p}. For $L=1$, the allowed $J^P$ assignments are $0^-$, $1^-$ and $2^-$, corresponding to the spectroscopic labels $ ^3 P_0$, $ ^3 P_1$ and $ ^3 P_2$, respectively. The discrepancies between \lq\lq CT" and \lq\lq CT+Ex." results for the channels with $(S,I)=(-1,0)$, $(-1,1)$, $(2,1/2)$, $(0,3/2)$ and $(1,1)$ are entirely attributable to the $u$-channel exchanges. The $s$-channel contributions to these channels, by contrast, are identical to zero, as they are prohibited by the conservation of additive quantum numbers like strangeness. Conversely, in the $(1,1)$ and $(0,1/2)$ channels, the $s$-channel exchanges of pseudoscalar and vector charmed mesons contribute significantly to the $P$-wave scattering lengths for the $0^-$ and $1^-$ cases, respectively. Additionally, the $u$-channel contribution is found to be sizeable for interactions involving a pion, such as $D^\ast\pi$ with $I=1/2,3/2$ and $D_s^\ast\pi$ with $I=1$. As seen from the scattering lengths in Tables~\ref{Tab:sl} and~\ref{tab:sl.p}, meson-exchange diagrams have a negligible effect on the $S$-wave scattering lengths, whereas they can considerably modify the $P$-wave ones. Accordingly, in what follows we confine our analysis to the $S$-wave sector and retain only the contact-term contribution.

\subsection{Pion mass dependence of the scattering lengths\label{sec.sl.mpi}}

Extensive lattice QCD studies of scatterings involving pNGBs and charm mesons already exist in the literature~\cite{Liu:2012zya,Mohler:2012na,Mohler:2013rwa,Moir:2016srx,Bali:2017pdv,Cheung:2020mql,Gayer:2021xzv,Yan:2024yuq,Lang:2022elg,Lang:2025pjq}. While the vast majority of these works have focused on the $\mathcal{P}\phi$ sector, investigations into the $\mathcal{P}^\ast\phi$ interactions have only recently become available for the $(S,I)=(0,1/2)$ channel~\cite{Lang:2022elg,Lang:2025pjq}. More simulations are certainly needed to fully understand the interactions between pNGBs and vector charm mesons, which are crucial for unveiling the nature of the $D_{s1}(2460)$ and other $1^+$ states. To this end, in this subsection we provide predictions for the $S$-wave scattering lengths at unphysical pion masses, which are suitable for direct comparison with existing and future lattice QCD results.

Given that the strange quark mass is fixed, the pion-mass dependence of the $K$, $D^\ast$ and $D_s^\ast$ masses can be derived from Eqs~\eqref{eq.pi.K.mass}, \eqref{eq.mass.Dst} and~\eqref{eq.mass.Dsst}, resulting in
\begin{align}
M_K^2={\mathring{M}_K^2+\frac12 M_\pi^2}\ ,   \quad
M_{D^{\ast}}^2={\mathring{M}_{D^{\ast}}^2+2(2h_0^\ast+h_1^\ast)M_\pi^2}\ , \quad
M_{D^{\ast}_s}^2=\mathring{M}_{D^{\ast}_s}^2+4h_0^\ast M_\pi^2\ ,
\label{eq:mass1}
\end{align}
where $\mathring{M}_K^2\equiv B m_s$, $\mathring{M}_{D^{\ast}}^2\equiv \overline{M}^{2}_{\mP^\ast}+4h_0^\ast \mathring{M}_K^2$ and $\mathring{M}_{D^{\ast}_s}^2\equiv \overline{M}^{2}_{\mP^\ast}+4(h_0^\ast+h_1^\ast) \mathring{M}_K^2$ are the masses of $K$, $D^\ast$ and $D^\ast_s$ mesons in the limit of $M_\pi \to 0$, or equivalently, $\hat{m}\to 0$. Their values are determined, by using the physical masses as inputs, to be: $\mathring{M}_{K}=486.3$~{\rm MeV}, $\mathring{M}_{D^{\ast}}=2003.5$~{\rm MeV},  $\mathring{M}_{D^{\ast}_{s}}=2111.6$~{\rm MeV}. In addition, we use Eq.~\eqref{eq.eta.etap} to extrapolate the $\eta/\eta^\prime$ masses and their mixing angle $\theta$, with $M_0=835.7$~MeV taken from Ref.~\cite{Guo:2015xva}.

\begin{figure*}[tbp]
\centering
\includegraphics[width=0.975\textwidth]{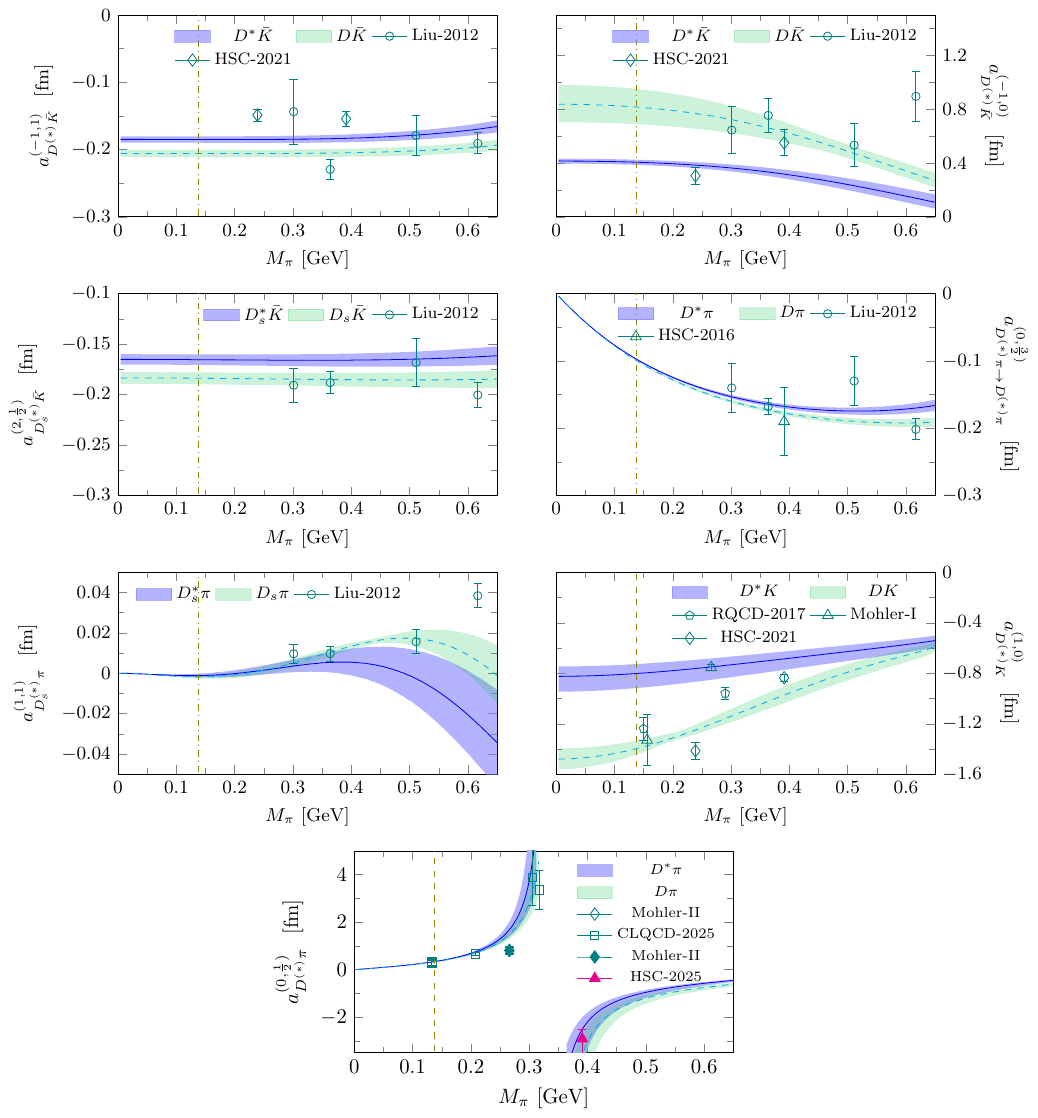} 
\caption{Pion-mass dependence of $S$-wave $\mathcal{P}^\ast\phi$ scattering lengths of elastic channels. The error bands are obtained by varying the NLO LECs within their $1$-$\sigma$ uncertainties. The olive dashed vertical line denote physical pion mass. For comparison, we also show lattice QCD data: Liu-2012~\cite{Liu:2012zya}, Mohler-I~\cite{Mohler:2013rwa}, Mohler-II~\cite{Mohler:2012na}, HSC-2016~\cite{Moir:2016srx}, RQCD-2017~\cite{Bali:2017pdv}, HSC-2021~\cite{Cheung:2020mql}, CLQCD-2025~\cite{Yan:2024yuq},
 HSC-2025~\cite{Lang:2025pjq}.
}
\label{fig:SL}
\end{figure*}

The predicted $S$-wave scattering lengths of $\mathcal{P}^\ast\phi$ interactions as functions of $M_{\pi}$ up to $650$~MeV are shown in Fig.~\ref{fig:SL}. Also displayed for comparison are the $S$-wave $\mathcal{P}\phi$ scattering lengths from Ref.~\cite{Guo:2015dha}, which clearly reveal the symmetry breaking effects caused by the differences of heavy meson masses. The figure shows that, except for the $D^{\ast}\pi$ channel with $I=1/2$, the scattering lengths in all other elastic channels generally increase or decrease to some extent without changing signs. A completely different trend is observed in the $(0,{1}/{2})$ channel, where the $D^{\ast} \pi$ scattering length grows rapidly toward positive infinity, then abruptly turns toward negative infinity. This behavior is closely related to the evolution of the $(0,{1}/{2})$ poles in the complex energy plane, to be discussed in detail in subsection~\ref{sec.pole.pion.mass}. The scattering length diverges to negative infinity, which is a clear signature of a strongly attractive interaction. This behavior typically indicates that the system forms a bound state close to the threshold. 

Fig.~\ref{fig:SL} also includes currently available lattice QCD data to facilitate direct comparison. The hollow and filled markers denote the existing $\mathcal{P}\phi$~\cite{Liu:2012zya,Mohler:2013rwa,Mohler:2012na,Moir:2016srx,Bali:2017pdv,Cheung:2020mql,Yan:2024yuq} and $\mathcal{P}^\ast\phi$~\cite{Mohler:2012na,Lang:2025pjq} data, respectively. Notably, our result for the $D^\ast\pi$ scattering length with $I=1/2$ is in excellent agreement with the most recent lattice QCD determination by the HSC~\cite{Lang:2025pjq}. The HSC value, $a\simeq -2.91_{-0.59}^{+0.39}$~fm at pion mass $M_\pi=391$~MeV, is indicated by the magenta triangle in the figure.\footnote{This value is extracted from the $K$ matrix parametrization of the charmed axial-vector $D^\ast\pi$-$D^\ast\eta$-$D_s^\ast\bar{K}$ amplitude given in that work. } This agreement confirms the reliability of the values of LECs obtained from HQSS, as presented in Table~\ref{value:LECs}.

\subsection{The $S$-matrix parameters}

\begin{figure}
    \centering
    \includegraphics[width=1.05\linewidth]{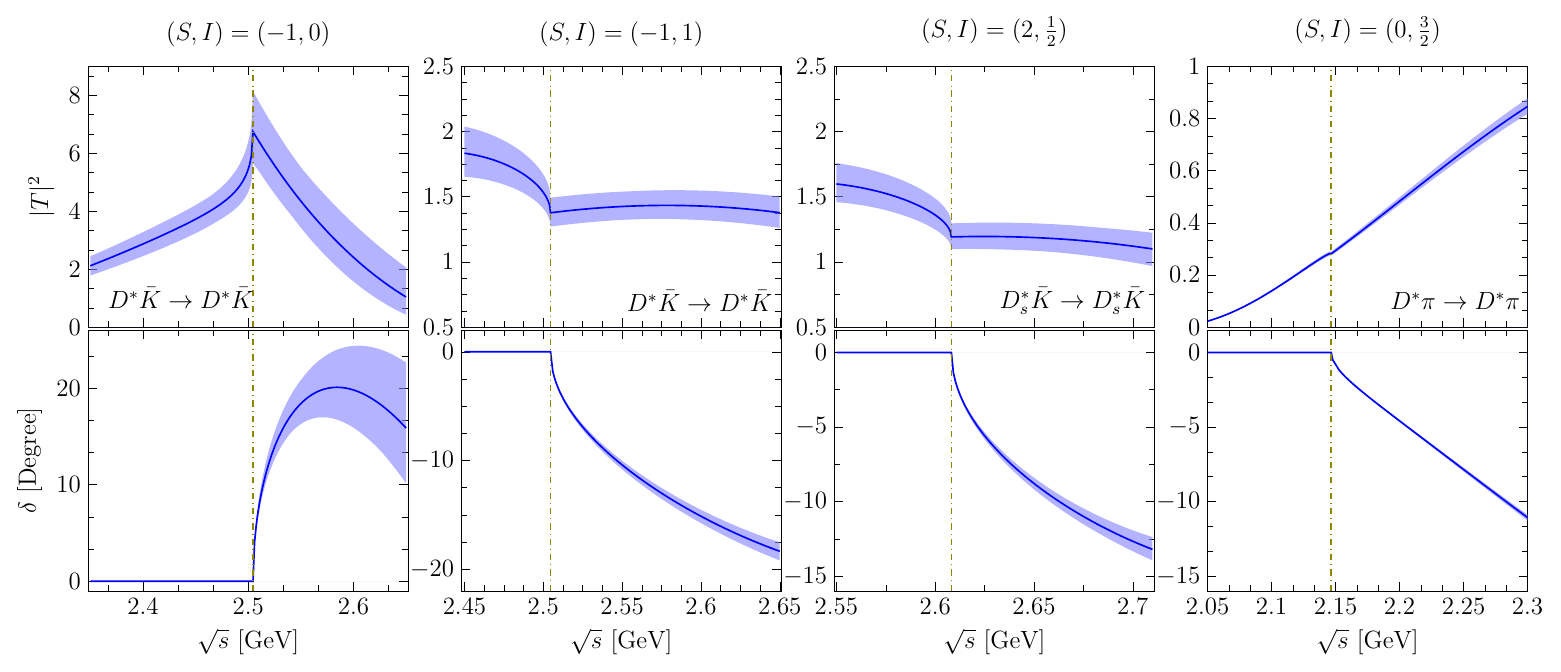}
    \caption{Moduli of amplitude squared and phase shifts for the four single channels at physical masses. The channels with $(S,I)=(-1,0)$, $(-1,1)$, $(2,1/2)$ and $(0,3/2)$ are shown from left to right.}
    \label{fig:amp2.ps.sc}
\end{figure}

In addition to the scattering lengths, physical observables such as phase shifts and inelasticity, collectively referred to as $S$-matrix parameters, can provide important information on the scattering amplitudes. The $S$-matrix is related to the $T$ matrix  via
\begin{align}
S_{ij}(s)=\delta_{ij}+2i\rho_i^{1/2}T_{ij}(s)\rho_j^{1/2} \ , \quad T_{ij}=-\frac{1}{16\pi} \mathcal{T}_{ij}\ ,
\end{align}
where $\rho_i(s)=\sigma_i(s)/s$, the subscripts $i,j$ label the corresponding channels. We stress that the $T$ matrix differs with the unitaried amplitude $\mathcal{T}$ (c.f., Eq.~\eqref{eq.BSE.matrix}) by an overall factor $-1/(16\pi)$. One can parametrize the diagonal matrix elements as
\begin{align}
S_{ii}=\eta_i e^{2i\delta_{i}}\ ,
\end{align}
where $\eta_i$ and $\delta_i$ are inelasticity and phase shift, respectively. Both of them are real numbers. In view of the unitary condition $(S^\dagger S)_{ii}=\sum_j|S_{ji}|^2=1$, one immediately has $0\leq\eta_i=|S_{ii}|\leq 1$. In practice, the inelasticity and phase shift can be conveniently extracted from the $T$ matrix using the following formulae
\begin{align}
\eta_i=|1+2i\rho_iT_{ii}|\ ,\quad \delta_i={\rm Arg}\bigg[\frac{1-\eta_i}{2i}+\rho_iT_{ii}\bigg]\ .
\end{align}

\begin{figure}
    \centering
    \includegraphics[width=0.9\linewidth]{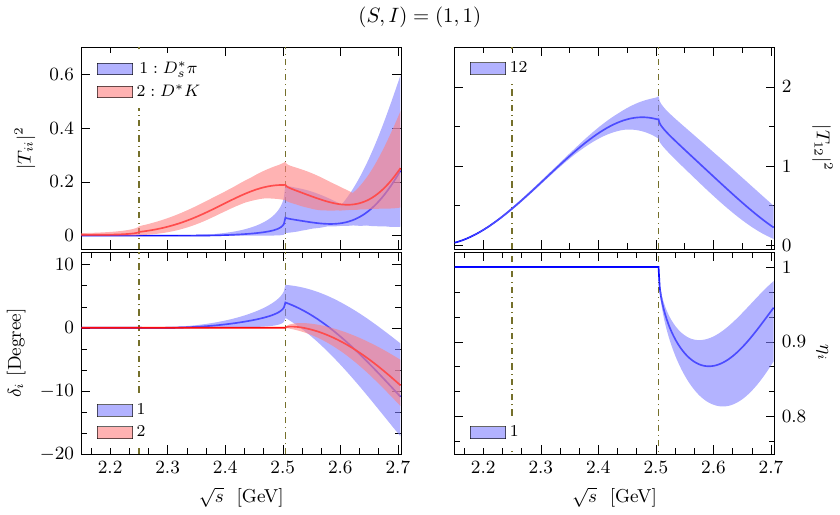}
    \caption{Moduli of amplitude squared, phase shifts and inelasticity for the $(1,1)$ coupled channels at physical masses.}
    \label{fig:amp2.11}
\end{figure}

\begin{figure}
    \centering
    \includegraphics[width=0.9\linewidth]{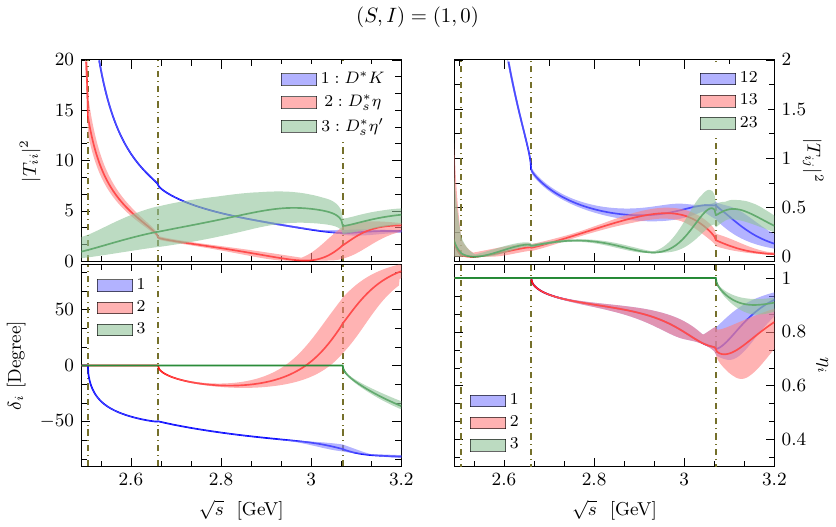}
    \caption{Moduli of amplitude squared, phase shifts and inelasticity for the $(1,0)$ coupled channels at physical masses.}
    \label{fig:amp2.10}
\end{figure}

\begin{figure}
    \centering
    \includegraphics[width=0.9\linewidth]{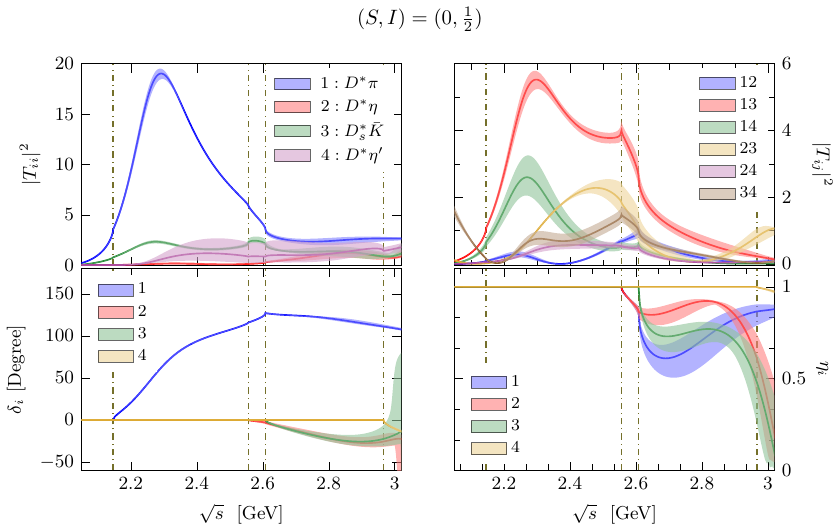}
    \caption{Moduli of amplitude squared, phase shifts and inelasticity for the $(0,1/2)$ coupled channels at physical masses.}
    \label{fig:amp2.012}
\end{figure}

The moduli of amplitude squared and phase shifts for the four single-channel processes are shown in Fig.~\ref{fig:amp2.ps.sc}. For the coupled channels, namely, $(1,1)$, $(1,0)$ and $(0,1/2)$, we present the moduli of amplitude squared, phase shifts and inelasticities in Figs.~\ref{fig:amp2.11}, \ref{fig:amp2.10} and~\ref{fig:amp2.012}, respectively. The uncertainties are propagated from the errors of LECs. Such quantities are essential for high-precision Dalitz plot analyses of $B$ meson decays, which can be measured at LHCb and Belle II. Moreover, the line shapes encode vital information about the underlying structure of the amplitudes, including direct manifestations of dynamically generated poles (see, e.g. Ref.~\cite{Zhang:2024qkg}). For instance, the leftmost panels of Fig.~\ref{fig:amp2.ps.sc} show a cusp structure at the $D^\ast\bar{K}$ ($I=0$) threshold, suggesting the presence of a possible subthreshold resonance. As another example, in Fig.~\ref{fig:amp2.012}, the broad bump of $T_{11}$ between the $D^\ast\pi$ and $D^\ast\eta$ thresholds, together with the sharp rise of the phase shift $\delta_1$ past 90 degrees, provide a strong indication of a near-threshold resonance. In the next section, we will perform a detailed pole analysis of the $\mathcal{P}^\ast\phi$ interactions and scrutinize the nature of the poles by studying their trajectories as functions of, e.g., the pion mass $M_\pi$ and $N_C$.

\section{Dynamically generated poles and their trajectories}\label{sec:pole_trajectories}

\subsection{Poles corresponding to positive-parity charmed mesons \label{sec.poles}}
According to the $S$-matrix theory, unstable resonances and stable bound states manifest as pole singularities of the unitarized amplitudes when analytically continued to the appropriate RS. In this section, we investigate the $1^+$ states that are dynamically generated in the $S$-wave charmed vector mesons and pNGBs scattering amplitudes, with the aim of elucidating their physical interpretation and inner structure. The various RSs can be classified by the signs of the imaginary part of $g_i(s)$ in Eq.~\eqref{eq.G.matrix}. Specifically, for $N$ coupled channels, an RS is labeled by an $N$-tuple $[\xi_1\xi_2\cdots\xi_N]$ with $\xi_i\in \{0,1\}$. In this labeling, $\xi_i=0$ means that the $g_i(s)$ is used, whereas $\xi_i=1$ indicates that the $g_i(s)$ is replaced by $g_i^{\rm II}(s)\equiv g_i(s)-2 {\rm Im}[g_i(s)]$. Consequently, the index $n$ of the $n$-th RS is determined by $n=1+\sum_{i=1}^N \xi_i 2^{i-1}$. For instance, for a three-coupled-channel case, the $1$st RS, also called the physical RS, corresponds to $[000]$, while the $2$nd and $3$rd RSs are denoted by $[100]$ and $[110]$ respectively. Furthermore, the $n$-th RS is often labeled using Roman numerals (cf. Table~\ref{Tab:poles}). In addition to the pole positions, the associated residues also possess direct physical interpretations. In the neighborhood of a given pole $s_{\rm pole}$, the unitarized amplitude $\mathcal{T}_{ij}$ admits a Laurent series as $\mathcal{T}_{ij} = {g_ig_j}/{(s-s_{\rm pole})}+\cdots$, where $g_{i}$ and $g_j$ denote its couplings to the $i$-th and $j$-th channels. 

\begin{table*}[!htbp]
\centering
\renewcommand{\arraystretch}{1.1}
\caption{Predicted poles and moduli of their couplings to the different channels. The pole positions are given as $M-i\,\Gamma/2$ in units of MeV, and the couplings are in units of GeV.}
\label{Tab:poles}
\begin{tabular} {c c c c c c c c }
\noalign{\global\arrayrulewidth=2pt}
\hline
\noalign{\global\arrayrulewidth=0.4pt}
\multicolumn{1}{c}{ \multirow{2}*{$(S,I)$} } & \multicolumn{1}{c}{ \multirow{2}*{RS} } & \multicolumn{1}{c}{ \multirow{2}*{poles \ ($M- i\,\dfrac{\Gamma}{2} $)}} &\multicolumn{4}{c}{ $|$Residues$|$~(\rm{GeV}) } \\
\cline{4-7}
\multicolumn{2}{c}{}& &$|g_{1}|$     &$|g_{2}|$&$|g_{3}|$&$|g_{4}|$&\\
\hline
$(-1,0)$      &$\rm{\RNum{2}}$   
&$2362.1^{+13.6}_{-10.2}-i\,111.0_{-26.7}^{+25.1}$ &$11.2^{+1.0}_{-0.6}$($D^\ast \bar{K}$)  &--  &-- &-- \\
$(1,1)$      &$\rm{\RNum{2}}$   &$2576.7^{+30.5}_{-20.4}-i\,323.7_{-5.4}^{+3.0} $    &$6.3^{+0.6}_{-0.4}$ ($D^\ast_s\pi$)     &$11.8^{+0.3}_{-0.3}$ ($D^\ast K$) &-- &--  \\
        &$\rm{\RNum{3}}$   &$2347.1^{+15.1}_{-11.5}-i\,227.4_{-13.3}^{+11.5} $
        &$7.1^{+0.3}_{-0.2}$($D^\ast_s\pi$)  &$5.6^{+0.2}_{-0.2}$  ($D^\ast K$) &-- &-- \\
$(1,0)$            &$\rm{\RNum{1}}$   &$2455.2^{+3.2}_{-2.7}$   &$10.6^{+0.2}_{-0.2}$($D^\ast K$) &$7.0^{+0.0}_{-0.0}$($D^\ast_s\eta$) & $0.6^{+0.6}_{-0.6}$($D^\ast_s\eta^{\prime}$) & --  \\
$(0,\frac12)$  &$\rm{\RNum{2}}$  &$2255.6^{+3.3}_{-2.8}-i\,112.5_{-2.9}^{+2.6}$ &$10.3^{+0.1}_{-0.1}$($D^\ast \pi$) & $2.0^{+0.3}_{-0.3}$($D^\ast \eta$) &$5.2^{+0.1}_{-0.1}$ ($D^\ast_s \bar{K}$) &$4.4^{+0.6}_{-0.6}$ ($D^\ast \eta^{\prime}$)     \\
         &$\rm{\RNum{3}}$  &$2558.1^{+31.0}_{-23.4}-i\,207.2^{+7.8}_{-8.4}$ & $5.8^{+0.1}_{-0.0}$ ($D^\ast \pi$) & $5.3^{+0.4}_{-0.3}$($D^\ast \eta$) & $12.2^{+0.2}_{-0.2}$ ($D^\ast_s \bar{K}$)  & $2.3^{+0.4}_{-0.3}$ ($D^\ast \eta^{\prime}$)  \\
& $\rm{\RNum{4}}$ & { $2307.1^{+16.1}_{-12.1} -i\,186.4^{+20.9}_{-14.5} $} & $7.4^{+0.3}_{-0.4}$($D^\ast\pi$)  &$5.0^{+0.2}_{-0.0}$($D^\ast\eta$)  &$5.0^{+0.7}_{-0.3}$($D^\ast_s\bar{K}$) 
&$4.7^{+2.3}_{-1.2}$($D^\ast\eta^\prime$)
         \\
\noalign{\global\arrayrulewidth=2pt}
\hline
\noalign{\global\arrayrulewidth=0.4pt}
\end{tabular}
\end{table*}

Our results of pole positions, quoted as $M-i\Gamma/2$, and moduli of the couplings $|g_{i}|$ are compiled in Table~\ref{Tab:poles}. For the $D^\ast \bar{K}$ single channel with $(S,I)=(-1,0)$, a resonance pole at $\sqrt{s_{\rm pole}}=(2362.1-i\,111.0)~\rm{MeV}$ is found in the second RS. As can be seen from Fig.~\ref{fig:amp2.ps.sc}, the pole would only induce a moderate threshold cusp structure in the $D^\ast\bar{K}$ invariant mass distribution, making it difficult to resolve experimentally. Its counterpart in the $\mathcal{P}\phi$ sector is a virtual state~\cite{Guo:2015dha}, which belongs to the scalar $\mathbf{6}$ multiplet and has been found on lattice in Ref.~\cite{Cheung:2020mql}. By HQSS, this state corresponds to the isoscalar member of the axial-vector sextet in the exact flavor SU(3) symmetry limit, which is a pure $\mathbf{6}$ state without any mixture from other multiplets ($\bar{\mathbf{3}}$ and $\overline{\mathbf{15}}$). We will verify this in next subsection by carrying out an SU(3) study.

For the $(S,I)=(1,1)$ case, we find two broad resonance poles. One pole is located above the $D^\ast K$ threshold on the second RS, while the other lies below the $D^\ast K$ threshold on the third RS. The former resides deep in the complex $\sqrt{s}$ plane and should be interpreted as the shadow pole~\cite{Eden:1964zz} of the latter. 

\begin{table*}[!htbp]
\centering
\caption{Comparison of our predictions with previous UChPT results taken from Refs.~\cite{Du:2017zvv,Guo:2018gyd}, along with the PDG averages~\cite{ParticleDataGroup:2026aaa}.}
\label{Tab:poles.comparison}
{
\begin{tabular} {c c c c c c c c c c}
\noalign{\global\arrayrulewidth=2pt}
\hline
\noalign{\global\arrayrulewidth=0.4pt}
$(S,I)$  & RS & poles \ ($M-i\,\frac{\Gamma}{2}$) & SU(3)-HQSS~\cite{Du:2017zvv} & Guo~\cite{Guo:2018gyd} & {PDG~\cite{ParticleDataGroup:2026aaa}}\\
\hline
$(1,0)$            &$\rm{\RNum{1}}$   &$2455.2^{+3.2}_{-2.7}$   & $2456^{+15}_{-21}$& $2431(36)$ &$2459.5 \pm 0.6 $  \\
$(0,\frac12)$  &$\rm{\RNum{2}}$  &$2255.6^{+3.3}_{-2.8}-i\,112.5_{-2.9}^{+2.6}$  &$2247^{+5}_{-6} - i\,107^{+11}_{-10} $ & $2228^{-8}_{+1}-i\,182^{+44}_{-28}$ &\multirow{2}{*}{$2412(9) + i\,314(29 ) $}   \\
         &$\rm{\RNum{3}}$  &$2558.1^{+31.0}_{-23.4}-i\,207.2_{-8.4}^{+7.8}$  & $2555^{+47}_{-30}-i\,203^{+8}_{-9}$  & $2606_{+23}^{-30}-i\,59^{+13}_{-25}$ & 
         \\
\noalign{\global\arrayrulewidth=2pt}
\hline
\noalign{\global\arrayrulewidth=0.4pt}
\end{tabular}
}
\end{table*}

For the coupled-channel scattering with $(S,I)=(1,0)$, a bound-state pole, $\sqrt{s_{\rm pole}}=2455.2^{+3.2}_{-2.7}~\rm{MeV}$, appears just below the $D^\ast K$ threshold and exhibits a strong coupling to the $D^\ast K$ channel. This pole can be identified with the $D_{s1}(2460)$ state, as its mass is consistent with the PDG average~\cite{ParticleDataGroup:2026aaa} for the $D_{s1}(2460)$ within $1$-$\sigma$ uncertainty (see table~\ref{Tab:poles.comparison}). In Table~\ref{Tab:poles.comparison}, a comparison is also made with the pole results obtained from the NLO SU(3) chiral potentials (see the fourth column), as reported in Ref.~\cite{Du:2017zvv}. Lattice QCD also provides a first-principle study of the charmed-strange state with high statistics~\cite{Bali:2017pdv}. The mass of the $D_{s1}(2460)$ obtained therein is $2451\pm4~\rm{MeV}$ at $M_{\pi}=150~\rm{MeV}$~\cite{Bali:2017pdv}, which is already close to the physical pion mass. Our prediction is in agreement with the lattice QCD determination. 

In the $(0,{1}/{2})$ channel, an intriguing phenomenon is the two-pole structure of the axial-vector charmed state $D_1(2430)$, analogous to that of its HQSS partner, the scalar charmed meson $D_0^\ast(2300)$. The two-pole structure of $D_0^\ast(2300)$ occurs naturally in the framework of UChPT, simultaneously implementing chiral symmetry, unitarity and coupled-channel effects, and has been extensively identified in many works~\cite{Albaladejo:2016lbb, Du:2017zvv, Guo:2018gyd, Asokan:2022usm, Zhuang:2026lta, Luo:2026kui}. For an instructive review on the two-pole structures in QCD, we refer the readers to Ref.~\cite{Meissner:2020khl}. Based on the current U(3) UChPT framework for the $\mathcal{P}^\ast \phi$ interactions, the resulting two $I(J^P)=\frac{1}{2}(1^+)$ poles are located at $2255.6^{+3.3}_{-2.8}-i\,112.5_{-2.9}^{+2.6}$ (RS II) and $2558.1^{+31.0}_{-23.4}-i\,207.2_{-8.4}^{+7.8}$ (RS III), respectively, which are in good agreement with the findings of Ref.~\cite{Du:2017zvv}. The relevant results from Ref.~\cite{Guo:2018gyd} are listed in the penultimate column of Table~\ref{Tab:poles.comparison}. Despite the large uncertainties, a noticeable difference remains in the higher pole. The lower pole lies approximately $110$~MeV above the $D^\ast \pi$ threshold and predominantly couples to this channel. The higher pole resides slightly above the $D^\ast\eta$ threshold, but exhibits its strongest coupling to the $D_s^\ast\bar{K}$ channel. In addition, a shadow pole emerges on the fourth RS, residing between the $D^\ast\pi$ and $D^\ast\eta$ threshold.

\subsection{SU(3) study}

\begin{figure}[ht]
    \centering
    \includegraphics[width=0.85\linewidth]{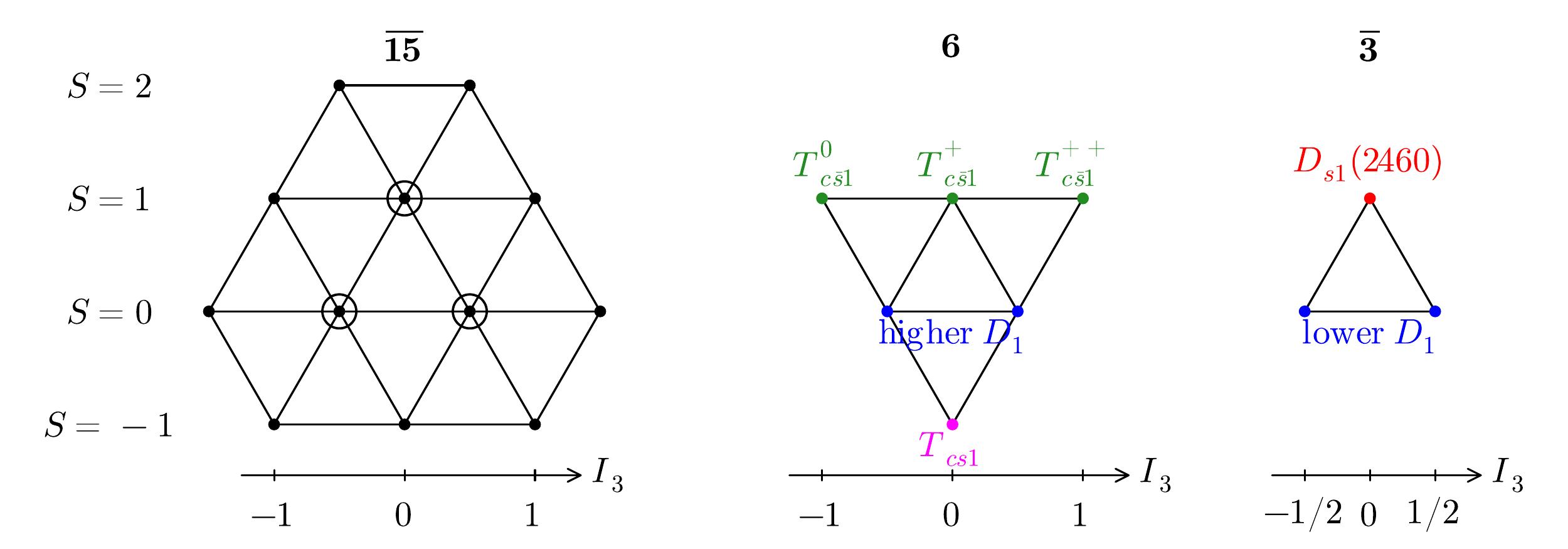}
    \caption{SU(3) weight diagrams of $\overline{\mathbf{15}}$, $\mathbf{6}$ and $\overline{\mathbf{3}}$ {\it irreps}.}
    \label{fig:weights}
\end{figure}

From the perspective of flavor SU(3) group, the vector charmed mesons ($c\bar{q}$ with $q=u,d,s$) form an anti-triplet {\it irrep} ($\overline{\mathbf{3}}$), whereas the pNGBs ($\pi$, $K$, $\bar{K}$ and $\eta_8$) belong to the octet {\it irrep} ($\mathbf{8}$). Their interactions can therefore be classified according to the direct-product reduction $\overline{\mathbf{3}}\otimes\mathbf{8}=\overline{\mathbf{15}}\oplus \mathbf{6}\oplus \overline{\mathbf{3}}$, whose weight diagrams are displayed in Fig.~\ref{fig:weights}. In practice, however, it is more appropriate to characterize the $\mathcal{P}^\ast\phi$ interactions using the subgroup $\textrm{SU(2)}\otimes \textrm{U(1)}\subset \textrm{SU(3)}$, labeled by isospin $I $ and strangeness $S$, respectively, owing to the hierarchy $m_u\simeq m_d \ll m_s$. In the following, we shall identify the poles found in various $(S,I)$ channels (see Table~\ref{Tab:poles}) with their counterparts in the SU(3) {\it irrep} classification. 

To that end, we restore the SU(3) symmetry by enforcing the degenerate quark-mass limit $m_u=m_d=m_s$. In this limit, the pNGB and charmed meson masses acquire the common values $\overline{M}_0$ and $\overline{M}_{\mathcal{P}^\ast}$, respectively. In our numerical computation, we use $\overline{M}_0=0.49$~GeV for the pNGBs and $\overline{M}_{\mathcal{P}^\ast}=2.04$~GeV for the vector charmed mesons. Following Ref.~\cite{Albaladejo:2016lbb}, we interpolate between the physical and SU(3) symmetric cases by continuously varying the meson masses via  
\begin{align}
M_{\phi}&= M_{\phi}^{\rm phy.} + 
x(\overline{M}_0-M_{\phi}^{\rm phy.}) \ ,\quad \phi\in\{\pi,K,\bar{K},\eta\}\ ,\notag\\
M_{\mathcal{P}^\ast}&=M_{{\mathcal{P}}^\ast}^{\rm phy.}
+x(\overline{M}_{\mathcal{P}^\ast} - M_{\mathcal{P}^\ast}^{\rm phy.} )\ , \quad \mathcal{P}^\ast\in\{D^\ast, D_s^\ast\}\ ,
\end{align}
with the physical masses taken from Table~\ref{cof:mass}. The interpolation parameter $x$ ranges from $0$ to $1$, where $x=0$ and $x=1$ correspond to the physical and SU(3) symmetric limits, respectively. A subtle issue arises regarding the number of RS. In the SU(3) limit, where all channels share a common threshold, only two RSs exist. However, for $x\neq 1$ with symmetry breaking, a $N$-coupled channel interaction gives rise to $2^N$ RSs, labeled by $[\xi_1\xi_2\cdots\xi_N]$. In Ref.~\cite{Albaladejo:2016lbb}, the $[\xi_1\xi_2\cdots\xi_N]$ RS is continued into either the $[00\cdots0]$ sheet or the $[11\cdots1]$ sheet by gradually changing $\xi_i$. As a result, the $[00\cdots0]$ and $[11\cdots1]$ sheets are identified with the physical and unphysical sheets, respectively, in the SU(3) limit. In this work, we do not adopt this procedure. Instead, we track the poles that can naturally move to the RSs defined in the SU(3) symmetric case.

\begin{figure}[ht]
    \centering
    \includegraphics[width=0.9\linewidth]{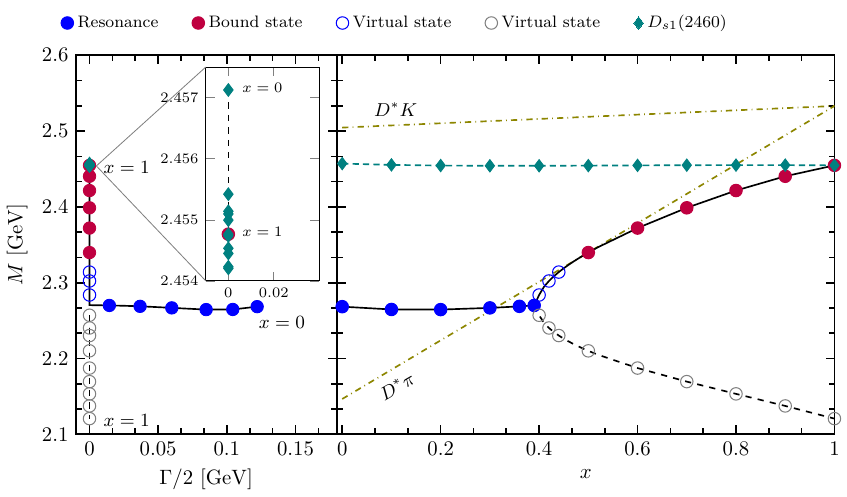}
    \caption{Evolution of the $(1,0)$ bound-state and the $(0,1/2)$ RS-II poles, with $\sqrt{s_{\rm pole}}=M-i\Gamma/2$, from the physical case ($x=0$) to the SU(3) symmetric limit ($x=1$). The circles and diamonds mark the lower $D_1(2430)$ and the $D_{s1}(2460)$ poles, respectively. At $x=1$, the poles merge into an SU(3) triplet pole. Left: pole trajectories in the complex plane. Right: real parts of pole positions as functions of $x$. The olive dash-dotted lines indicate the relevant thresholds.}
    \label{fig:triplet}
\end{figure}

Fig.~\ref{fig:triplet} shows the evolution of the $D_{s1}(2460)$ and the lower $D_1(2430)$ poles with respect to $x$. The $D_{s1}(2460)$ pole remains on the first RS and gradually evolves into a bound state belonging to the $\overline{\mathbf{3}}$ {\it irrep} in the SU(3) symmetric limit. The lower $D_1(2430)$ pole and its conjugate partner traverse the $D^\ast\pi$ threshold on RS-II, move downwards the real axis, and eventually coalesce into a pair of virtual states. One of the virtual states (indicated by the blue circles in the figure) then approaches to the threshold, transforms into a bound-state pole (red dots) on the first RS and ultimately merges with the $D_{s1}(2460)$ pole at $x=1$. Thus, the isoscalar $D_{s1}(2460)$ and the lower isodoublet $D_1(2430)$ poles can be identified as flavor SU(3) partners, together forming a triplet under SU(3).

\begin{figure}[ht]
    \centering
    \includegraphics[width=0.85\linewidth]{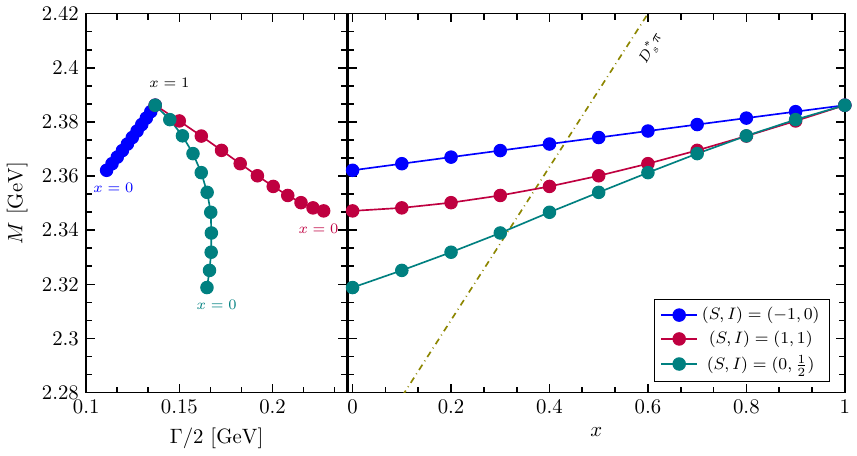}
    \caption{Evolution of the $(-1,0)$ RS-II, the $(1,1)$ RS-III and the $(0,1/2)$ RS-IV poles (see legend for color coding), with $\sqrt{s_{\rm pole}}=M-i\Gamma/2$, from the physical case ($x=0$) to the SU(3) symmetric limit ($x=1$). At $x=1$, the poles merge into an SU(3) sextet pole. Left: pole trajectories in the complex plane. Right: real parts of pole positions as functions of $x$. The olive dash-dotted line indicates the $D_s^\ast\pi$ threshold.}
    \label{fig:sextet}
\end{figure}

Analogously, Fig.~\ref{fig:sextet} displays the paths followed by the $(-1,0)$ RS-II, the $(1,1)$ RS-III and the $(0,1/2)$ RS-IV poles. The trajectories of the higher $D_1(2430)$ pole at $\sqrt{s}=2558.1-i\,207.2$ MeV and the $(1,1)$ RS-II pole at $\sqrt{s}=2576.7-i\,323.7$~MeV, are omitted, as their evolution is taken over by their corresponding shadow poles, i.e., the $(1,1)$ RS-III and the $(0,1/2)$ RS-IV poles.\footnote{It was first proposed in Ref.~\cite{Eden:1964zz} that resonances correspond not to a single pole, but to a dominant pole with a series of shadow poles on different RSs. Therein, it is also argued that the evolution of the dominant pole can be taken over by one of the shadow poles as the symmetry breaking effect is gradually switched off.}  As the parameter $x$ increases, the poles converge to the same point $\sqrt{s}=2386.2-i\,137.0$~MeV on RS-II, thereby constituting a sextet under SU(3). One finds that the sextet pole resides below the degenerate $\mathcal{P}^\ast\phi$ threshold (c.f. the $D^\ast_s\pi$ threshold in the figure), and hence corresponds to a subthreshold resonance in the ideal case of exact SU(3) symmetry. Being located below the lowest relevant threshold, this subthreshold resonance does not manifest as a BW peak in the physical scattering amplitude. Instead, it appears as a virtual-state-like enhancement in the near-threshold region, and can only be accessed indirectly through coupled-channel effects. The poles in the physical situation, with SU(3) breaking turned on, inherit this common feature, reflecting the persistent weakness of the attraction in the $\mathbf{6}$ {\it irrep} channel.

\subsection{Pole trajectory with varying pion mass\label{sec.pole.pion.mass}}

\begin{figure}
    \centering
    \includegraphics[width=0.6\linewidth]{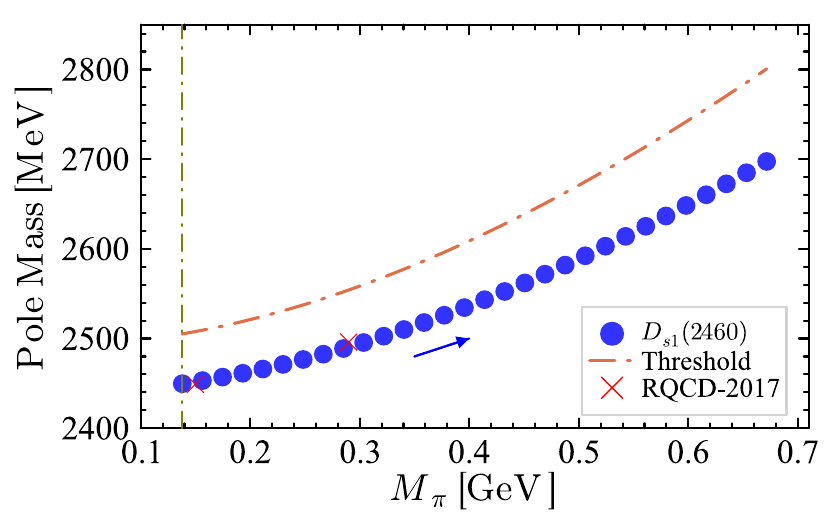}
    \caption{Pole trajectory of the $(1,0)$ bound state as a functions of $M_\pi$. The lattice QCD data are taken from Ref.~\cite{Bali:2017pdv} and shown as red crosses for easy comparison.}
    \label{fig:mpi_S1I0}
\end{figure}

In this part, we investigate the trajectories of the poles as the pion mass $M_{\pi}$ varies, up to approximately $5 M_\pi^{\rm phy.}$. The pion mass dependence formulae for the pNGB masses, the charmed vector masses, the $\eta$-$\eta^\prime$ mixing angle are the same as the ones used in subsection~\ref{sec.sl.mpi}. For the $D_{s1}(2460)$ state in the $(1,0)$ channel, its evolution with respect to $M_{\pi}$ is illustrated in Fig.~\ref{fig:mpi_S1I0}. This state remains a bound state below the $D^{*}K$ threshold throughout the entire considered $M_\pi$ region, with its binding energy increasing slightly. The RQCD results for the $D_{s1}$ state, extracted at two different unphysical pion masses $M_\pi=150$ and $290$~MeV but in the same finite volume $L/a=64$ (see Table III of Ref.~\cite{Bali:2017pdv}), are shown as red crosses in the figure for comparison, and a good agreement is found.

\begin{figure}
    \centering
    \includegraphics[width=0.6\linewidth]{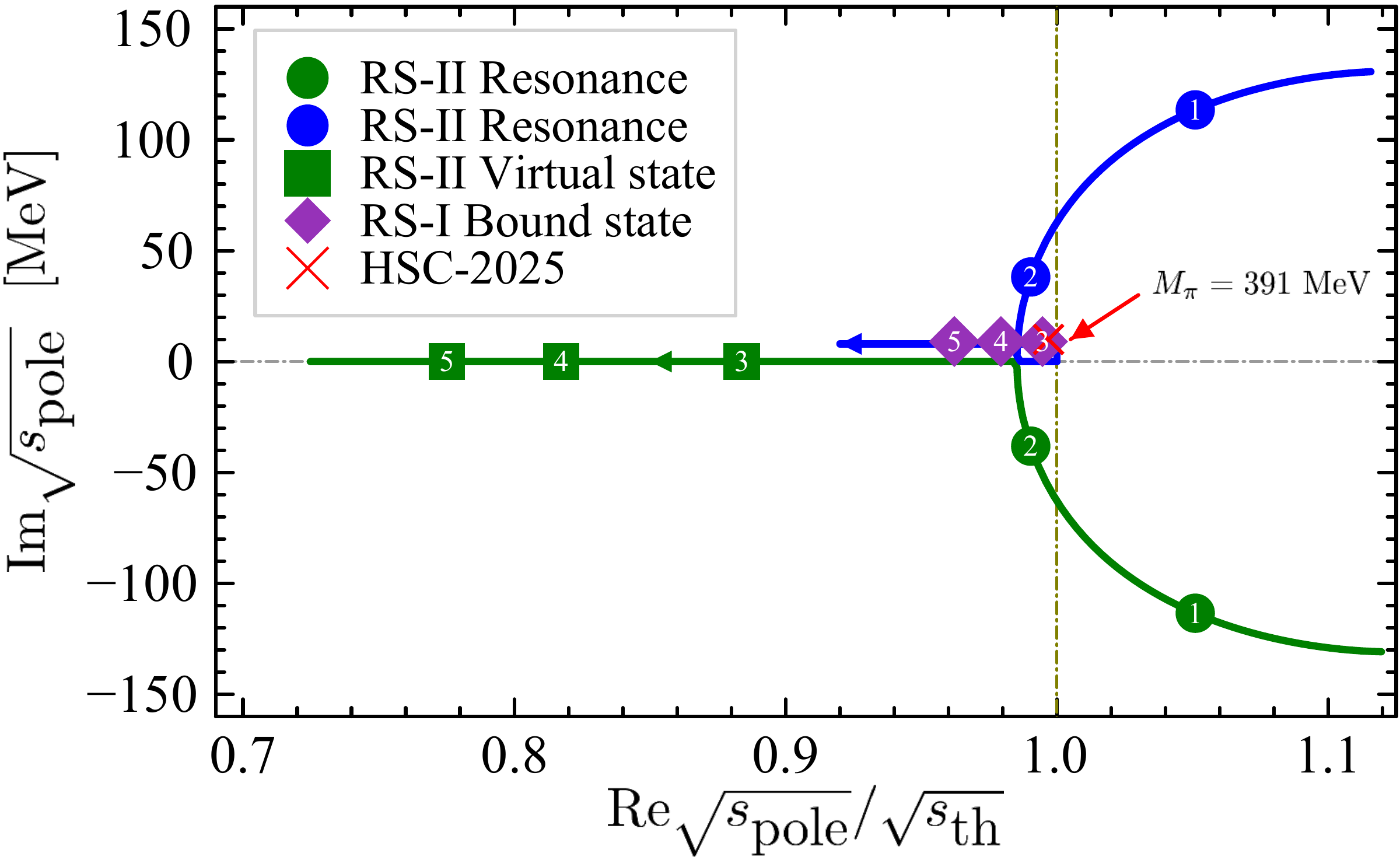}
    \caption{Pole trajectory of the $(0,1/2)$ RS-II resonance as a function of $M_\pi$. The integers $n$ denote the multiples of the physical pion mass, i.e., $M_\pi=n M_\pi^{\rm phy.}$. The red cross represents the lattice QCD result by the Hadron spectrum collaboration~\cite{Lang:2025pjq}: $\sqrt{s}_{\rm pole}^{\rm HSC}=2395.6\pm1.4$~MeV at $M_\pi= 391$~MeV. The olive dash-dotted vertical line denotes the $D^\ast\pi$ threshold.}
    \label{fig:mpi_S0I12_RS2_low}
\end{figure}

The trajectory of the lower pole in the $(0,{1}/{2})$ channel, which lies below the $D^{*}\eta$ threshold, follows a more exotic pattern. As shown in Fig.~\ref{fig:mpi_S0I12_RS2_low}, this pole (blue dots) and its conjugation (green dots) keep as a pair of resonances in the complex energy plane for $M_\pi< 288.8$~MeV. At $M_\pi = 288.8$~MeV, the pair of resonance states falls onto the real axis below the $D^{*}\pi$ threshold on RS-II and becomes a pair of virtual states. One virtual state (green diamonds) runs along the negative real axis toward negative infinity, while the other moves toward the threshold, hits the $D^{*}\pi$ threshold at $M_\pi = 338$~MeV, and becomes a bound state (magenta diamonds) on the first RS. The existence of a bound state for $M_\pi > 338$~MeV is also supported by a recent result from the HSC~\cite{Lang:2025pjq}, indicated by the red cross in the figure. Namely, based on finite volume energy levels from lattice QCD simulation performed at $M_\pi = 391$~MeV, they identified an axial-vector bound state with $\sqrt{s_{\rm pole}}= (2395.6\pm1.4)$~MeV just below the $D^{*}\pi$ threshold. The $M_\pi$ behavior of the lower pole can be understood as follows. As $M_\pi$ increases, the $D^\ast\pi$ interaction of derivative form, proportional to pion momentum and $M_\pi$, becomes more attractive. Since the $D^\ast\pi$ interaction dominates the formation of the lower $D_1(2430)$ pole, its increasing strength ultimately drives the resonance state into a bound state.

\begin{figure}
    \centering
    \includegraphics[width=0.6\linewidth]{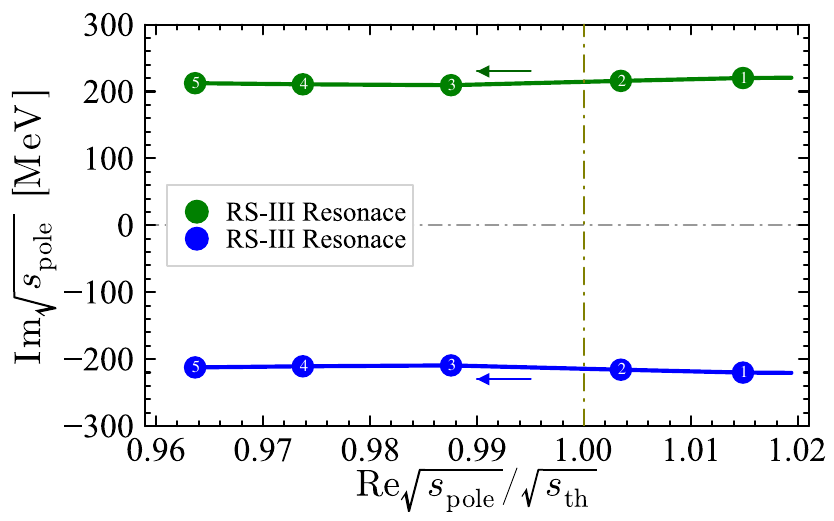}
    \caption{Pole trajectory of the $(0,\frac12)$ RS-III resonance as a function of $M_\pi$. The numbers within the circular markers indicate the multiples of the physical pion mass. The olive dash-dotted vertical line denotes the $D^\ast\eta$ threshold.}
    \label{fig:mpi_S0I12_RS2_high}
\end{figure}

On the other hand, the broad resonance on RS-$\rm{\RNum{3}}$ in the $(0,1/2)$ channel moves along a simple trajectory, persisting as a resonance with a nearly constant half-width, as shown in Fig.~\ref{fig:mpi_S0I12_RS2_high}. For $M_\pi=391$~MeV, the pole position from our study is $\sqrt{s_{\rm pole}}=2592.8^{+31.0}_{-23.4}-i\,209.6^{+8.4}_{-7.8}$~MeV, which is $26$~MeV below the $D^\ast\eta$ threshold. Compared to the HSC lattice QCD result of $\sqrt{s_{\rm pole}}=2737(79)-i\,221(88)$~MeV~\cite{Lang:2025pjq}, the imaginary parts agree well within uncertainties, whereas our real part is about 144 MeV lower than the lattice simulation. It was pointed out in Ref.~\cite{Asokan:2022usm} that, concerning the two-pole structure of $D_0^\ast(2300)$, the higher pole relies strongly on the parametrization adopted in fitting to lattice QCD data~\cite{Moir:2016srx}. Specifically, the higher pole, which is lost in the traditional $K$ matrix analysis~\cite{Moir:2016srx}, emerges again when an improved $K$ matrix incorporating SU(3) symmetry constraints is utilized. However, even with this improved treatment, a higher pole consistent with our UChPT prediction has not yet been established~\cite{Lang:2025pjq}. A further refined $K$-matrix approach incorporating correct chiral behavior, as proposed in Ref.~\cite{Du:2025beb}, may hopefully provide a final resolution to this issue.

\subsection{Pole trajectory with varying $N_C$}

The large $N_C$ behaviors of the dynamically generated poles encode information on their inner structure. For instance, if the $N_C$ trajectory of a pole on the energy plane approaches real axis when $N_C\to \infty$, the pole is most likely a BW resonance with a dominant $\bar{q}q$ component~\cite{Pelaez:2003dy,Xiao:2005rg,Pelaez:2006nj,Guo:2021blc}. To that end, the $N_C$ counting of the relevant parameters in the pNGBs-vector-charmed-meson scattering amplitudes are assigned as follows. First, the mass of any $\bar{q}q$ meson scales as a constant when $N_C$ varies, as pointed out in Refs.~\cite{tHooft:1973alw,Witten:1979kh}. Therefore, the $N_C$ counting for the meson masses, $M_\pi$, $M_K$, $\overline{M}_{\mathcal{P}^\ast}$, are of $\mathcal{O}(1)$. Second, the $N_C$ scaling of the pNGB decay constant constant $F_0$ in the chiral limit is $\mathcal{O}(\sqrt{N_C})$~\cite{Coleman:1980mx,Veneziano:1979ec}. Third, the singlet $\eta_0$ mass squared, $M_0^2$, scales as $\mathcal{O}(1/N_C)$, due to the fact that the QCD U(1)$_A$ is responsible for the massive $\eta_0$ and is of $\mathcal{O}(1/N_C)$~\cite{Witten:1979vv,Veneziano:1979ec,Coleman:1980mx,Manohar:1998xv}. Fourth, the $N_C$ counting rule of the NLO $\mathcal{P}^\ast\phi$ LECs $h_{i=0,\cdots,5}^\ast$ can be assigned analogously to those in $\mathcal{P}\phi$ scattering~\cite{Guo:2009ct,Guo:2015dha}. Each trace in the chiral operators leads to one more power of $1/N_C$; see e.g. Refs.~\cite{Manohar:1998xv,Pelaez:2004xp} for more details.  
By counting the number of flavor traces in the LO and NLO Lagrangians, c.f. Eqs~\eqref{lolag:ct} and~\eqref{nlolag}, one straightforwardly obtains $h_{1,3,5}^\ast\sim  \mathcal{O}(1)$ and $h_{0,2,4}^\ast \sim \mathcal{O}(1/N_C)$. Finally, it is natural to assume the $N_C$ scaling of the subtraction constant $a_i(\mu)$ in Eq.~\eqref{eq:g} to be $\mathcal{O}(1)$ in the large $N_C$ expansion, as illustrated in Ref.~\cite{Guo:2012yt}. Note that in this work we confine ourselves to the leading $N_C$ counting of the parameters.

\begin{figure}
    \centering
    \includegraphics[width=0.6\textwidth]{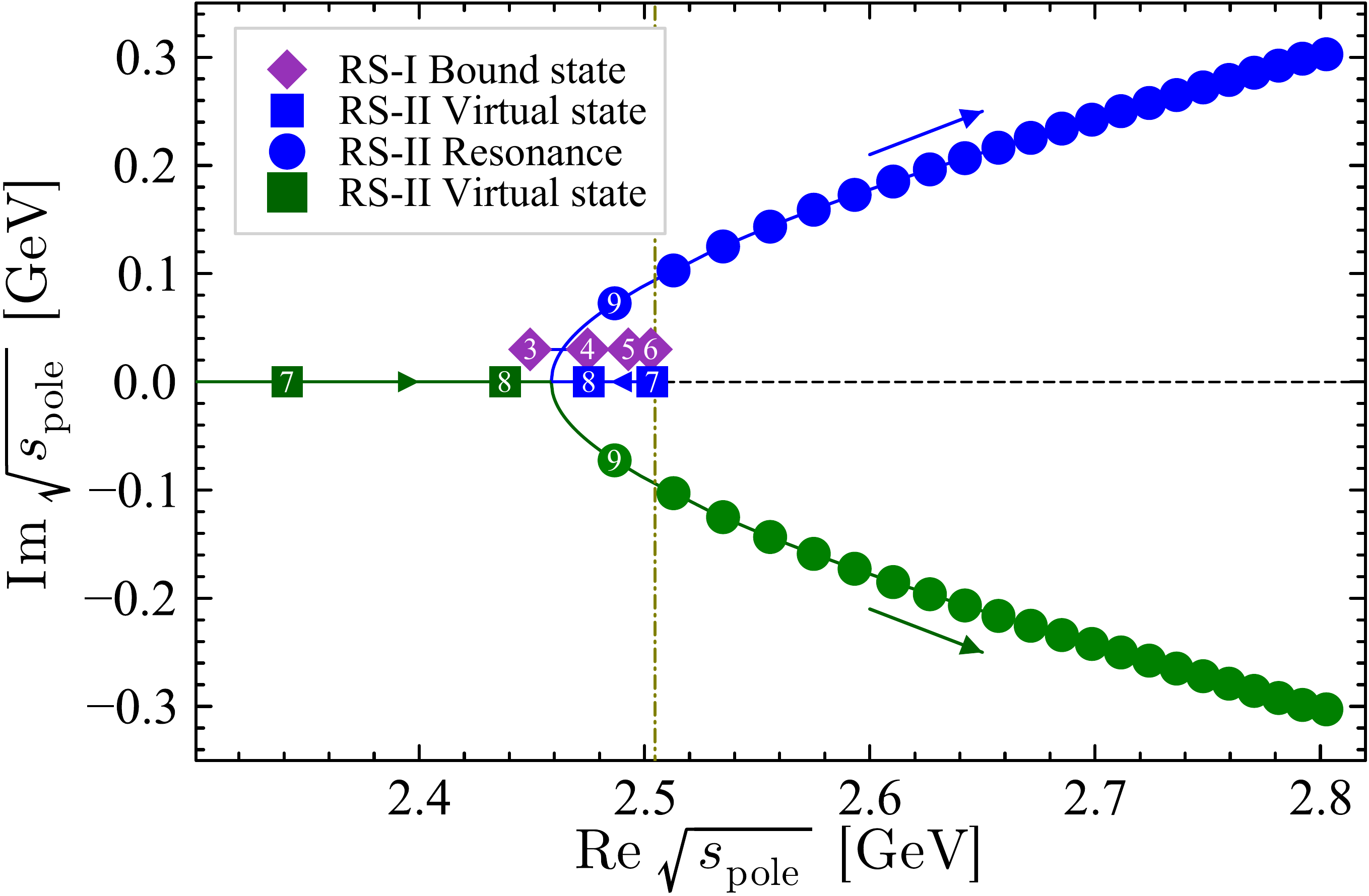}
    \caption{Pole trajectory of the $(1,0)$ pole a function of $N_C$. The bound, virtual, and resonance states are represented by diamonds, squares and circles, respectively. The physical case corresponds to $N_C=3$. The olive dash-dotted vertical line denotes the $D^\ast K$ threshold.}
    \label{fig:Ds12460.NC}
\end{figure}

With the above $N_C$ counting rules of all the involved parameters, the $N_{C}$ dependence of $D_{s1}(2460)$ pole can be obtained, which is shown in Fig.~\ref{fig:Ds12460.NC}. As $N_C$ increases, the bound-state pole approaches towards the $D^\ast K$ threshold, hits the branch point at around $N_C\simeq7$ and converts into a virtual-state pole on the RS-II. This pole collides with another pole, moving from negative real infinity, and becomes a pair of conjugated resonance poles. These two poles move to infinity on the complex plane, rather than falling on to the positive real axis, indicating the $D_{s1}(2460)$ is not of $\bar{q}q$ nature. 

\begin{figure}
    \centering
    \includegraphics[width=0.975\textwidth]{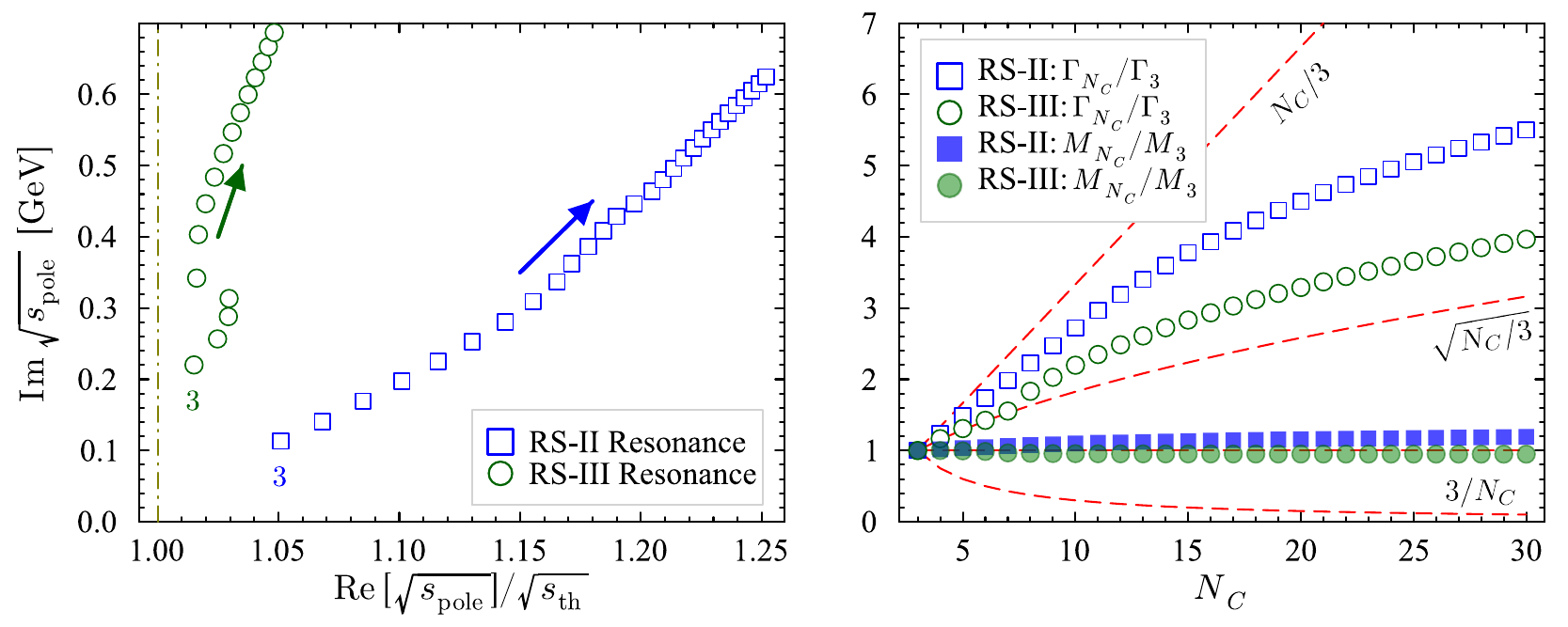}
    \caption{Left panel: pole trajectory of the $(0,1/2)$ resonances as a function of $N_C$ in the complex plane. The RS-II and RS-III poles are represented by blue squares and green circles, respectively. The physical case corresponds to $N_C=3$. The olive dash-dotted vertical line denotes the $D^\ast\pi$ ($D^\ast\eta$) threshold for the RS-II (RS-III) pole. Right panel: mass and width $N_C$ behavior of the $(0,1/2)$ resonances, normalized to their respective values at $N_C=3$, denoted by $M_3$ and $\Gamma_3$. Three typical $N_C$ scaling rules, $N_C$, $\sqrt{N_C}$ and $1/N_C$ are shown as red dashed lines for reference.}
    \label{fig:S0I12_pole.NC}
\end{figure}

The RS-II and RS-III poles in the $(0,1/2)$ channel are also numerically traced. The evolution of the pole positions in the complex energy plane are shown in left panel of Fig.~\ref{fig:S0I12_pole.NC}, where the leftmost points correspond to the physical case with $N_C=3$. Analogous to the $D_{s1}(2460)$ pole, both of the $(0,1/2)$ poles go to complex infinity as $N_C\to \infty$. The right panel of Fig.~\ref{fig:S0I12_pole.NC} shows the mass $M$ and width $\Gamma$, defined via $\sqrt{s_{\rm pole}}=M-i\Gamma/2$, as functions of $N_C$, normalized respectively to their $N_C=3$ values. For both poles, the masses scale as $M_{\rm pole}\sim N_C^0$, while the widths follow $\Gamma_{\rm pole}\propto N_C^\alpha$ with ${1}/{2}\leq\alpha<1$ rather than $\alpha=-1$. This $N_C$ pattern for the axial-vector charmed resonances, analogous to that observed for the scalar $f_0(500)$ and $K^\ast_0(700)$ resonances in Ref.~\cite{Pelaez:2003dy,Guo:2011pa}, is in conflict with a $\bar{q}q$ interpretation while suggesting possible tetraquark or molecular nature.

\section{Summary}
\label{sec:summary}

With the aim of clarifying the nature of axial-vector charmed mesons, $\mathcal{P}^\ast\phi$ scattering has been studied in a relativistic U(3) chiral effective field theory at NLO. The obtained scattering amplitudes are organized into four elastic and three coupled channels according to strangeness $S$ and isospin $I$. The BSE under on-shell approximation is employed to construct the unitarized amplitudes. The NLO LECs showing up in the $\mathcal{P}^\ast\phi$ potentials are estimated by HQSS through the well-established $\mathcal{P}\phi$ LECs. Consequently, both $S$- and $P$-wave scattering lengths are predicted. The meson-exchange contributions are found to be negligible for the former but sizable in the $P$-wave channels with $(S,I)=(1,0)$ and $(0,1/2)$. We show the $M_\pi$ dependence of the $S$-wave scattering lengths, which facilitates direct comparison with lattice QCD at large unphysical pion masses. Phase shifts together with inelasticities are presented for all elastic channels as well.

Particular emphasis is placed on dynamically generated $S$-wave $J^P=1^+$ states, above all the $D_{s1}(2460)$ and $D_{1}(2430)$. Poles are extracted by analytical continuation of the unitarized amplitudes on the appropriate RSs. In the $(-1,0)$ channel, a broad resonance is found. The $(1,1)$ coupled-channel amplitude accommodates two broad resonances, located above and below the $D^\ast K$ threshold on RS-II and RS-III, respectively. In the $(1,0)$ channel, a bound-state pole appears just below $D^\ast K$ threshold at $\sqrt{s_{\rm pole}}\simeq 2455$~MeV and is naturally identified with $D_{s1}(2460)$, in agreement with previous SU(3) ChPT determinations~\cite{Du:2017zvv,Guo:2018gyd} and lattice QCD result~\cite{Bali:2017pdv}. In the $(0,1/2)$ channel, a characteristic two-pole structure emerges, with the experimentally-observed $D_{1}(2430)$ interpreted as their interference. 

To further unveil the nature of $D_{s1}(2460)$ and $D_{1}(2430)$, their pole trajectories are examined as functions of $M_\pi$, $N_C$, and the SU(3)-symmetry-breaking parameter $x$. In the SU(3) flavor-symmetric limit, $D_{s1}(2460)$ and the lower $D_1(2430)$ pole belong to the same flavor SU(3) $\bar{\mathbf{3}}$ {\it irrep}, whereas the higher $D_1(2430)$ pole originates from the $\mathbf{6}$ {\it irrep}. Upon increasing $M_\pi$, the bound-state character of $D_{s1}(2460)$ and the resonance nature of the higher $D_1(2430)$ pole persist, while the lower $D_1(2430)$ pole evolves from a resonance into a bound state---a trend compatible with recent HSC lattice QCD finding~\cite{Lang:2025pjq}. In the large-$N_C$ limit, all these poles move to complex infinity, disfavoring a conventional $\bar{q}q$ assignment and supporting a hadronic-molecule interpretation.

The $\mathcal{P}^\ast\phi$ scattering amplitudes obtained in the present work provide inputs for high-precision Dalitz plot analyses of $B$ meson decays at LHCb and Belle~II, and are likewise essential for the description of final-state interaction phenomena. Future advances will require more precise lattice QCD and experimental data in more channels.

\acknowledgments 

We would like to thank Ling-Yun~Dai, Feng-Kun~Guo, Liuming~Liu, Haobo~Yan and Han-Qing~Zheng for helpful discussions. This work is supported by Hebei Natural Science Foundation under Grant No.~A2025205018; by Science Research Project of Hebei Education Department under Contract No.~QN2025063; by National Nature Science Foundations of China (NSFC) under Contract No.~12547166, No.~12275076, No.~12335002, No.~12475078; by the Science Fund for Distinguished Young Scholars of Hunan Province under Grant No.~2024JJ2007; by the Fundamental Research Funds for the Central Universities under Contract No. 531118010379; by the Science Foundation of Hebei Normal University with Contract No.~L2025B09 and No.~L2023B09.

\appendix

\section{Explicit expressions for the coefficients abbreviated in Table~\ref{tab:ciex}} \label{Apend:Co}

In this appendix, we show the coefficients that are not given explicitly in Table~\ref{tab:ciex}. For the $D^\ast K$-$D_s^\ast \eta$-$D_s^\ast \eta^\prime$ coupled channels with $(S,I)=(1,0)$, the relevant coefficient read 
\begin{align}
\mathcal{C}^{(1,0) \ K\eta}_1&=\frac{-M_K^2(5c_\theta+4\sqrt{2}s_\theta)+3M_{\pi}^2c_\theta}{2\sqrt{3}} \ , \\
\mathcal{C}^{(1,0) \  \eta\eta}_0&=
\frac13 \big[c_\theta^2(4M_K^2-M_{\pi}^2)+4\sqrt{2}c_\theta s_\theta(M_K^2-M_\pi^2)
+s_\theta^2(2M_K^2+M_\pi^2)\big] \ ,\\
\mathcal{C}^{(1,0) \  \eta\eta}_1&=\frac{2(M_{\pi}^2-2M_K^2)(\sqrt{2}c_\theta+s_\theta)^2}{3} \ ,\\
\mathcal{C}^{(1,0) \  K\eta^\prime}_1&=\frac{M_K^2(4\sqrt{2}c_\theta-5 s_\theta)+3M_\pi^2 s_\theta}{2\sqrt{3}} \ ,\\
\mathcal{C}^{(1,0) \  K\eta^\prime}_{35}&=\frac{s_\theta-2\sqrt{2}c_\theta}{\sqrt{3}} \ ,\\
\mathcal{C}_S^{(1,0) \  \eta\eta^\prime}&=-2(\sqrt{2}c_\theta^2-c_\theta s_\theta-\sqrt{2}s_\theta^2) \ ,\\
\mathcal{C}_U^{(1,0) \  \eta\eta^\prime}&=-2(\sqrt{2}c_\theta^2-c_\theta s_\theta-\sqrt{2}s_\theta^2) \ ,\\
\mathcal{C}^{(1,0) \  \eta\eta^\prime}_0&=
\frac{2(M_\pi^2-M_K^2)(\sqrt{2}c_\theta^2-c_\theta s_\theta-\sqrt{2}s_\theta^2)}{3} \ ,\\
\mathcal{C}^{(1,0) \ \eta\eta^\prime}_1&=\frac{2(2M_K^2-M_\pi^2)(\sqrt{2}c_\theta^2-c_\theta s_\theta-\sqrt{2}s_\theta^2)}{3}  \ ,\\
\mathcal{C}^{(1,0) \ \eta\eta^\prime}_{35}&=\frac{-2(\sqrt{2}c_\theta^2-c_\theta s_\theta-\sqrt{2}s_\theta^2)}{3} \ , \\
\mathcal{C}^{(1,0) \  \eta^\prime \eta^\prime}_0&=\frac13[s_\theta^2(4M_K^2-M_\pi^2)+4\sqrt{2}c_\theta s_\theta(M_\pi^2-M_K^2)
+c_\theta^2(2M_K^2+M_\pi^2)] \ , \\
\mathcal{C}^{(1,0) \ \eta^\prime \eta^\prime}_1&=\frac{2(M_\pi^2-2M_K^2)(\sqrt{2}s_\theta-c_\theta)^2}{3} \ , \\
\mathcal{C}^{(1,0)\ \eta^\prime \eta^\prime}_{35}&=\frac{2(\sqrt{2}s_\theta-c_\theta)^2}{3} \ .
\end{align}
For the $D^\ast\pi$-$D^\ast\eta$-$D_s^\ast\bar{K}$-$D^\ast\eta^\prime$ coupled channel with $(S, I)=(0,1/2)$, they are
\begin{align}
\mathcal{C}^{(0,\frac12)\  \eta \eta}_0&=\frac13\big[c_\theta^2(4M_K^2-M_\pi^2)+4\sqrt{2}c_\theta s_\theta(M_K^2-M_\pi^2)
+s_\theta^2(2M_K^2+M_\pi^2)\big] \ , \\
\mathcal{C}^{(0,\frac12)\  \eta \eta}_1&=\frac{-M_\pi^2(\sqrt{2}s_\theta-c_\theta)^2}{3} \ , \\
\mathcal{C}^{(0,\frac12)\  \eta \eta}_{35}&=\frac{(\sqrt{2}s_\theta-c_\theta)^2}{3} \ , \\
\mathcal{C}^{(0,\frac12)\  \bar{K}\eta}_1&=\frac{c_\theta(5M_K^2-3M_\pi^2)+4\sqrt{2}s_\theta M_K^2}{2\sqrt{6}} \ , \\
\mathcal{C}^{(0,\frac12)\  \bar{K}\eta}_{35}&=\frac{-(2\sqrt{2}s_\theta+c_\theta)}{\sqrt{6}} \ , \\
\mathcal{C}^{(0,\frac12)\  \eta\eta^\prime}_{(c_\theta,s_\theta)}&=\sqrt{2}c_\theta^2-c_\theta s_\theta-\sqrt{2}s_\theta^2 \ , \\
\mathcal{C}^{(0,\frac12)\  \eta\eta^\prime}_0&=\frac{2(M_\pi^2-M_K^2)(\sqrt{2}c_\theta^2-c_\theta s_\theta-\sqrt{2}s_\theta^2)}{3} \ , \\
\mathcal{C}^{(0,\frac12)\  \eta\eta^\prime}_1&=\frac{M_\pi^2(-\sqrt{2}c_\theta^2+c_\theta s_\theta+\sqrt{2}s_\theta^2)}{3} \ , \\
\mathcal{C}^{(0,\frac12)\  \eta\eta^\prime}_{35}&=\frac{\sqrt{2}c_\theta^2-c_\theta s_\theta-\sqrt{2}s_\theta^2}{3} \ , \\
\mathcal{C}^{(0,\frac12)\  \bar{K}\eta^\prime}_1&=\frac{(5M_K^2-3M_\pi^2)s_\theta-4\sqrt{2}M_K^2c_\theta}{2\sqrt{6}}\ , \\
\mathcal{C}^{(0,\frac12)\  \bar{K}\eta^\prime}_{35}&=\frac{2\sqrt{2}c_\theta-s_\theta}{\sqrt{6}} \ , \\
\mathcal{C}^{(0,\frac12)\  \eta^\prime\eta^\prime}_0&=\frac13\big[s_\theta^2(4M_K^2-M_\pi^2)+4\sqrt{2}c_\theta s_\theta(M_\pi^2-M_K^2)
+c_\theta^2(2M_K^2+M_\pi^2)\big]\ ,\\
\mathcal{C}^{(0,\frac12)\  \eta^\prime\eta^\prime}_1&=\frac{-M_\pi^2(\sqrt{2}c_\theta+s_\theta)^2}{3} \ , \\
\mathcal{C}^{(0,\frac12)\  \eta^\prime\eta^\prime}_{35}&=\frac{(\sqrt{2}c_\theta+s_\theta)^2}{3} \ .
\end{align}

\section{Explicit expressions of the Lorentz invariant amplitudes}\label{Apend:amp}
In this appendix, the chiral expressions of the Lorentz invariant functions $V_{i=1,\cdots,5}(s,t)$, defined in Eq.~\eqref{eq:Lorentzdec}, are shown explicitly. For brevity, we denote the contribution originating from the contact terms by $V_{\rm c}(s,t)$, whose expression reads
\begin{align}
V_{\rm c}(s,t)&=\frac{\mathcal{C}_{\rm LO}}{4F_0^2}(\Sigma-2s-t)+\frac{4h^\ast_0\mathcal{C}_0}{F_0^2}-\frac{2h^\ast_1\mathcal{C}_1}{F^2_0}
+\frac{(2h^\ast_2\mathcal{C}_{24}-h^\ast_3 \mathcal{C}_{35})(m_2^2+m_4^2-t)}{F_0^2}\notag\\
&+\frac{h^\ast_4\mathcal{C}_{24}-h^\ast_5 \mathcal{C}_{35}}{2F_0^2} \big[m_3^2m_4^2+m^2_1(m_{2}^2+2 m_{3}^2+m_{4}^2)
+m_{2}^2(m_{3}^2+2m_{4}^2) 
\notag\\
&+2s^2+2st+t^2-(2s+t)\Sigma \big]\ ,
\end{align}
where the abbreviation $\Sigma\equiv m_1^2+m_2^2+m_3^2+m_4^2$ is the sum of mass squared.

The amplitudes $V_1(s,t)$ and $V_2(s,t)$ include contributions from both contact diagrams and exchange diagrams. For $V_1(s,t)$, one has
\begin{align}
V_1(s,t)=
V_c(s,t)
&+\frac{g_1^2}{12F_0^2} \bigg[
\frac{\mathcal{C}_{S}}{M_{h^\ast}^2-s} \big[\Delta_{12}\Delta_{34}+s^2
+s(2t-\Sigma) \big] \notag\\
&+\frac{\mathcal{C}_U}{(\Sigma-M_{h^\ast}^2-s-t)}
\big[t^2+s \Sigma
-s^2-(\Sigma-2m_{2}^2) t 
+\Delta_{23}\Delta_{14}\big] \bigg] \ , \label{eq.inv.amp.V1}
\end{align}
with $\Delta_{ij}\equiv  m_i^2-m_j^2$.
For $V_2(s,t)$, one has
\begin{align}
V_2(s,t)&=V_c(s,t) + \frac{g_1^2}{24F_0^2} \frac{\mathcal{C}_U}{s(\Sigma-M_{h^\ast}^2-s-t)}
\bigg[m_1^2
\big[m_2^2(\Delta_{34}-s)-(m_4^2+s)(m_4^2-2s-t) 
\notag\\
&+m_3^2(m_4^2+6s-t)\big] 
-(m_2^2+m_4^2
-2s-t)
[m_2^2(\Delta_{34}+s)+s(\Delta_{34}-s-2t)] \bigg] 
\notag\\
&+\frac{g_0^2}{12F_0^2} \bigg[ 
\frac{\mathcal{C}_U}{\Sigma-M_{h}^2-s-t}
\big(\Sigma-2(m_{1}^2+m_{3}^2)
+s+\frac{\Delta_{12}\Delta_{34}}{s}\big)
+{\frac{4s \mathcal{C}_S}{s-M_{h}^2}}  \bigg]\ .\label{eq.inv.amp.V2}
\end{align}
At NLO, the $V_{i=3,4,5}(s,t)$ functions stem solely from the exchange diagrams, which are given by
\begin{align}
V_3(s,t)&=-\frac{g_0^2}{6F_0^2}\frac{\mathcal{C}_U}{\sqrt{s}(\Sigma-M_{h}^2-s-t)}(\Delta_{12}-s)+\frac{g_1^2}{12F_0^2}\frac{\mathcal{C}_U}{\sqrt{s}(\Sigma-M_{h^\ast}^2-s-t)}\notag\\
&\times
\bigg[(s-m_2^2)(m_2^2+m_4^2-2s-t)+m_1^2(m_2^2+m_4^2+2s-t)\bigg]\ ,\label{eq.inv.amp.V3}\\
V_4(s,t)&=-\frac{g_0^2}{6F_0^2}\frac{\mathcal{C}_U}{\sqrt{s}(\Sigma-M_h^2-s-t)}(\Delta_{34}-s)
+\frac{g_1^2}{12F_0^2}\frac{\mathcal{C}_{U}}{\sqrt{s}(\Sigma-M_{h^\ast}^2-s-t)}
\notag\\
&\times\bigg[m_2^2(\Delta_{34}+s)-(m_4^2-s)(m_4^2-2s-t)+m_3^2(m_4^2+2s-t)\bigg]\ ,\label{eq.inv.amp.V4}\\
V_5(s,t)&=\frac{g_0^2}{3F_0^2}\frac{\mathcal{C}_{U}}{\Sigma-M_{h}^2-s-t}+\frac{g^2_1}{6F_0^2}\bigg[\frac{2s\mathcal{C}_S}{M_{h^\ast}^2-s}
-\frac{\mathcal{C}_U(m_2^2+m_4^2-t)}{\Sigma-M_{h^\ast}^2-s-t}\bigg]\ .\label{eq.inv.amp.V5}
\end{align}

It is worth noting that, when exchange diagrams are neglected, the above expressions are simplified to
\begin{align}
V_{1}(s,t) = V_{2}(s,t) =V_c(s,t)\ , \quad
V_{3}(s,t) = V_{4}(s,t) = V_{5}(s,t)=0 \ .
\end{align}

\bibliography{Dstpi}

\end{document}